**Changes in seismicity in a volcanically active region of the Izu Peninsula, Japan**

K. Z. Nanjo,[a,b,c,d,*], Y. Yukutake[e], T. Kumazawa[c]

[a]*Global Center for Asian and Regional Research, University of Shizuoka, 3-6-1 Takajo, Aoi-ku, Shizuoka 420-0839, Japan*

[b]*Center for Integrated Research and Education of Natural Hazards, Shizuoka University, 836 Oya, Suruga-ku, Shizuoka 422-8529, Japan*

[c]*Institute of Statistical Mathematics, 10-3 Midori-cho, Tachikawa, Tokyo 190-8562, Japan*

[d]*Japan Agency for Marine-Earth Science and Technology, Yokohama Institute for Earth Sciences, 3173-25 Showa-machi, Kanazawa-ku, Yokohama, Kanagawa 236-0001, Japan*

[e]*Earthquake Research Institute, The University of Tokyo, 1-1-1 Yayoi, Bunkyo-ku, Tokyo 113-0032, Japan*

*Correspondence and requests for materials should be addressed to K.Z.N. (email: nanjo@u-shizuoka-ken.ac.jp)

ORCID: 0000-0003-2867-9185 (K. Z. N.); 0000-0002-1533-4885 (Y. Y.); 0000-0003-2435-640X (T. K.)

**Highlights**

- Deep low-frequency earthquakes in Izu Peninsula, Japan, showed seismic quiescence
- Shallow ordinary earthquakes in the same region showed similar quiescence
- The start of quiescence was earlier for shallow earthquakes than for deep ones
- Seismic quiescence occurred with no significant uplift during the study period
- We propose a hypothesis of magma behavior related to quiescence and no uplift

ABSTRACT

The eastern Izu Peninsula in Japan is volcanically and seismically active. Ordinary earthquakes frequently occurred at shallow depths in 2006 and 2009, when they clustered as swarms. Beneath ordinary earthquakes, low-frequency earthquakes (LFEs) were infrequently observed. To better understand the characteristics of those LFEs, we produced a LFE catalog for 2005-2020, using the matched-filter method. Timeseries analyses based on the Epidemic-Type Aftershock Sequence model showed quiescence, i.e., a change in occurrence rate, which became quiet. For comparison, the same analysis was conducted using the Japan Meteorological Agency catalog of ordinary earthquakes, and similar results were shown. The change points for both types of earthquakes fell during and after each of the swarms, revealing an earlier start of quiescence for shallow ordinary earthquakes than for deep LFEs. Surface displacement data obtained from the Geospatial Information Authority of Japan showed that the uplift was in abatement or unobservable during the study period. Our seismicity and geodetic results are interpretatively hypothesized as being associated with magmatic activity, as follows. When the absence of a supply of magma from a depth toward the crustal magma reservoir caused no significant uplift, magma intruded from this reservoir into the shallow crust, resulting in the 2006 and 2009 swarms. Due to this intrusion, stresses decreased in and around shallow parts of the reservoir and then quiescence for ordinary earthquakes began during or after each swarm. Due to this decrease, stresses in and around deep parts of it late decreased, delaying the start of quiescence of deep LFEs.

1. **Introduction**

   The Izu Peninsula in central Japan lies in the northern most part of the Izu Bonin Mariana (IBM) Arc and is located in the collision zone with Honshu Island, where crustal deformation is active (Fig. 1). This Peninsula is volcanically and seismically active (Aramaki and Hamuno, 1977). The eastern side of the Izu Peninsula is a monogenic volcano field, and its volcanoes are included in the Izu-Tobu Volcano Group (Koyama and Umino, 1991). This Volcano Group is listed as an active volcano where the Japan Meteorological Agency (JMA) operates 24-hours monitoring in real time. In the 1920s-1930s and 1970s-1990s, earthquake swarms, which consist of seismicity that lacks an obvious mainshock-aftershock sequence, and crustal deformation actively often occurred, and these were associated with magmatic activities (Geospatial Information Authority of Japan, 2016; Shishikura et al., 2023). Hereafter, the Geospatial Information Authority of Japan is referred to as GSI.

   A brief background summary of studies on swarms and magmatic activities on the eastern side of the Izu Peninsula is provided next. Ishida (1984) investigated a migrating swarm, which triggered the 1980 Izu-Hanto-Toho-Oki earthquake of magnitude $M$=6.7. Shimazaki (1988) suggested a dyke intrusion mechanism, and Tada and Hashimoto (1989) proposed a model of tensile fault. In 1989, an eruption occurred after the start of swarms, then formed a submarine volcano known as the Teishi Knoll (e.g., Oshima et al., 1991; Yamamoto et al., 1991; Nagamune et al., 1992), which is one of the members of the Izu-Tobu Volcano Group. The red triangle indicates the location of this volcano (Fig. 1). Geodetic and seismic data have been analyzed to explain these swarms on the basis of dike intrusion (Okada and Yamamoto, 1991; Okada et al., 2000; Hayashi and Morita, 2003; Morita et al., 2006; Miyamura et al., 2010). The swarm activity in 1998 was effectively modelled by the dike opening process (Okada et al., 2000). The swarm's hypocenters in 1998 were mainly aligned on a thin vertical plane (Hayashi and Morita, 2003; Morita et al., 2006), and the normal direction to the plane strongly coincided with the direction of tectonic extensional stress (Ukawa, 1991). Miyamura et al. (2010) and the Japanese governmental Earthquake Research Committee (ERC) revealed that the duration of earthquake swarms was associated with volumetric strain records at the Higashi-Izu JMA station (Fig. 1), and ERC (2010) forecasted the magnitudes of swarm activities. Kumazawa et al. (2016) used the Epidemic-Type Aftershock Sequence (ETAS) model (Ogata, 1988, 1992; Ogata and Tsuruoka, 2016) to assess eight swarms after mid-1980 and volumetric strain records at the Higashi-Izu JMA station, suggesting that the ETAS model may be helpful in monitoring magma intrusions that drive changes in stress.

In volcanic areas, it cannot be assumed that an earthquake is a high-frequency volcanic earthquake. This is because there are a myriad of earthquake types occurring as swarms that are both proximal and distal to volcanos and related to magmatic processes. Two of the first papers written on the subject of the myriad of earthquake types at volcanoes are associated with Mt. Usu in Japan (Omori, 1911; Murakami, 1951). There are many other more modern articles and reviews on this subject (e.g., White et al., 1996; Nishimura et al., 2000; Nakamichi et al., 2003; Nakamichi et al., 2004; Power et al., 2004; Smith et al., 2009; Nicols et al., 2011).

The JMA distinguishes ordinary earthquakes and LFEs in and around Japan. Each event in the JMA catalog is classified based on subsidiary information, such as natural (ordinary) earthquakes, LFEs, artificial events, and others. LFEs are defined as earthquakes with dominant frequencies of 2-8 Hz (Moriwaki, 2017). There are three classes of LFEs in Japan: LFEs that concentrate beneath active volcanoes, those that exist along the boundary between the Philippine Sea Plate and the continental plate in western Japan (Obara, 2002), and those that form several isolated clusters in the intraplate regions (Aso et al., 2013). Thus, these are distinct phenomena with different causes. LFEs in our study region were those that occurred at a volcano, but not in a subduction zone, and also not in a region with an isolated cluster. Events classified by JMA as LFEs were defined as LFEs in this study. Similarly, events that were classified by the JMA as ordinary earthquakes with a higher frequency than dominant frequencies (2-8 Hz) of LFEs, irrespective of their tectonic or volcanic origin, were defined as ordinary earthquakes in this study.

On the eastern side of the Izu Peninsula, ordinary earthquakes (black symbols in Fig. 1) mainly occurred at depth of <20 km during the period 2005-2020 when the JMA catalog was used. Although black symbols can also be observed at a depth range of >20 km, according to the event classification of JMA, these earthquakes are ordinary earthquakes. LFEs occurred at a depth range of 30-40 km beneath ordinary earthquakes during the same time period (red symbols in Fig. 1). Since Abe et al. (2023), who conducted a receiver function analysis, showed that the crust is >35 km thick beneath the Izu Peninsula, LFEs are considered to occur around the crust-mantle boundary or in deep parts of the crust. The swarms on the eastern side of the Izu Peninsula, investigated by previous studies briefly summarized above, consisted of ordinary earthquakes according to the JMA catalog.

While previous studies focused on swarms and ordinary earthquakes, to our knowledge, no research was conducted on LFEs. The primary purpose of this study is to clarify the characteristics of LFEs. To contribute toward achieving this purpose, we studied LFEs using the ETAS model, and investigated whether LFEs can be modeled by ETAS statistics. We were interested in appreciating if there is a relationship between LFEs and ordinary earthquakes, to understand whether either or both of these are related to magmatic activity.

**2. Data**

*2.1. LFEs and ordinary earthquakes in the JMA catalogs*

Using the JMA earthquake catalog, more than 10,000 ordinary earthquakes with $M≥-1$ occurred during 2005-2020 in the region shown by Fig. 1. JMA defined nine swarms associated with the Izu-Tobu Volcano Group during 2005-2020 (JMA, 2014). Two out of the nine swarms were defined as major swarms (Fig. 2a) and the durations of the major swarms were defined as periods from Apr. 17, 2006 to May 12, 2006 and from Dec. 16, 2009 to Jan. 12, 2010 (JMA, 2014). Supplementary Fig. 1a indicates how seismicity changed with time, using several colors. Seismicity outside the Izu-Tobu region occurred irrespective of time (Supplementary Fig. 1b-e). On the other hand, during the two periods when the swarms starting in 2006 and 2009 occurred, active seismicity was detected in this region (Supplementary Fig. 1b,c). Supplementary Fig. 2 confirms that ordinary earthquakes in the former period were clustered with respect to time and space. The locations of earthquakes in the first 5 days (Supplementary Fig. 2c) seem to cover the spatial extent of earthquakes during the entire period. A similar feature was observed for earthquakes in the latter period (Supplementary Fig. 3). The locations of earthquakes with the sources of the two swarms modeled by GSI (2007) and JMA (2010) indicate a NW-SE trending alignment (Figs. 1b and 2c and Supplementary Figs. 1-3). The two major swarms defined by JMA (2014) were considered in this study. Our criteria to define swarms with respect to time and space were as follows. First, the time periods of the two swarms used in this study were the same as those defined by JMA (2014). Second, the spatial extent of each swarm was defined by the locations of earthquakes in the first 5 days since the start of the corresponding swarm. The locations of these swarms overlapped or were close to each other (Fig. 2c). Their depths ranged from 0 to 20 km (Fig. 1c). This is consistent with a previous swarm that started in 1989 (Hayashi and Morita, 2003; Morita et al., 2006).

LFEs with $M≥-1$ were observed 47 times during 2005-2020 in the region shown in Fig. 1, in which events classified by JMA as LFEs were plotted. During the periods of the two major swarms defined by JMA (2014), there was no clear clustering (Fig. 2b).

JMA, which employs conventional event-detection methods, has difficulty in detecting LFEs that are easily buried in noise due to their low signal-to-noise ratios. Given that LFEs in the Izu-Tobu region are deeper than ordinary earthquakes, it is likely difficult to detect LFEs with conventional event-detection methods employed by JMA. If a LFE catalog is produced by using the matched-filter (MF) method, which cross-correlates a template to continuous seismic signals (Yukutake, 2017;

Yukutake et al., 2019), it could resolve this difficulty, and be used to find LFEs.

As described below, we produced a LFE catalog using the MF method. Our resultant catalog included 895 LFEs during the 2005-2020 period. This is about 19 times more than the number of earthquakes (47) reported by JMA during the same period. This means that LFEs occurred more frequently than were previously thought.

Our study region, indicated by a black rectangle in Fig. 1b, is referred to in this paper as the Izu-Tobu region. This region was defined to include the main activity of two earthquake swarms in 2006 and 2009. Most LFEs during 2005-2020 in the JMA catalog were located in the Izu-Tobu region (Fig. 1).

*2.2. The MF method for LFEs*

The MF method has been widely used (Shelly et al., 2007; Peng and Zhao, 2009; Kato et al., 2012; Ross et al., 2019). In this study, the MF system used for detecting LFEs beneath the Hakone volcano in Japan (Yukutake, 2017; Yukutake et al., 2019) was modified so that it was applicable to the Izu-Tobu region. The cross-correlation coefficient (*CC*) used for the MF method in the present study and our previous studies (Yukutake, 2017; Yukutake et al., 2019) is the same as that used by Shelly et al. (2007) and Peng and Zhao (2009). Waveforms of continuous signals that were used in this study covered the Jan. 2005-Dec. 2020 period, as recorded by 20 seismic stations (Fig. 3a) with a three-component velocity seismometer in and around the Izu-Tobu region.

To prepare template LFEs, we used the JMA catalog to select events classified as LFEs in the Izu-Tobu region. This study relied on statistical analyses of the LFE catalog, which covered the studied time interval. It should be noted that the catalog may be controlled by the selection of template earthquakes in the MF analysis. Relatively large LFEs with $M \geq 0.2$ were selected in order to allow template waveforms to include more information on signals than on noise. Then, among them, LFEs that were recorded by at least six stations with a minimum signal-to-noise-ratio of 2 were selected (Yukutake et al., 2019).

The MF procedure to identify LFEs, briefly described in the following paragraphs, is the same as that of Yukutake (2017) and Yukutake et al. (2019). In this procedure, waveforms of already-identified LFEs ($M \geq 0.2$) listed in the JMA catalog were used as template waveforms. The MF method was then applied to continuous seismic signals. This approach allowed us to identify previously-undetected LFEs that were not included in the JMA catalog. This does not imply that detected events in the present study were initially classified as ordinary earthquakes. Only after our analysis were detected events reclassified as LFEs.

Three-component waveform records for each template LFE were used, applying a six-second time window beginning two seconds before the onset times of the theoretical S-wave arrivals. Both template waveforms and continuous waveforms were bandpass-filtered for 1-6 Hz and decimated at 20 Hz to reduce the calculation cost. This band was selected according to Yukutake (2017) and Yukutake et al. (2019), although other studies used a slightly narrower band, such as 1-4 Hz by Kurihara and Obara (2021). The $CC$ between a template and continuous waveform at each sampling time for every component at each station was calculated. This calculation was the same as that employed by Yukutake (2017) and Yukutake et al. (2019). After subtracting the hypocenter-to-station travel time of the theoretical S-wave, the time sequences of the correlation function throughout all channels were stacked. If the peak of the stacked correlation function exceeded a threshold level of nine times the median absolute deviation, then an event was identified as a candidate LFE. After multiple counts were removed, the location of the candidate was assigned to the hypocenter of the matched template LFE determined by JMA. Following Yukutake (2017) and Yukutake et al. (2019), magnitude was determined as the mean of the maximum amplitude ratios of the template with respect to the candidate. The MF procedure described above was applied to all waveform records in the period Jan. 2005-Dec. 2020, and a preliminary catalog, including candidate LFEs, was created, although LFEs identified by five or less stations were not included in this catalog.

*2.3. LFE catalog*

Less reliable LFEs were removed from the preliminary catalog to create a finalized catalog, as follows. Among candidate LFEs, false detection occasionally occurred due to contamination by other seismic signals such as teleseismic earthquakes. This contamination led to the detection of LFEs with a large $M$, so we visually inspected each template LFE and examined whether this was used to detect many candidate LFEs with $M$>1.5, a magnitude above which LFEs have never been recorded by JMA in the Izu-Tobu region since 2005. We considered that such template LFEs had a feature similar to teleseismic earthquakes and decided to eliminate them from the list of template LFEs. Thus, candidate LFEs detected by using the eliminated template LFEs were removed from the preliminary catalog, resulting in the intermediate catalog that included 2502 LFEs. Despite this quality test, an additional test was conducted, as described below.

The $CC$-values of LFEs in the intermediate catalog (Fig. 3b) ranged between 0.1 (poor correlation with a template LFE) and 1 (strong correlation with, and identical to, the corresponding template LFE). Setting the minimum $CC$ to a low value implies the use of an incomplete catalog influenced by the nature of low signal-to-noise ratios of LFEs. The minimum threshold for $CC$ ($CC_{th}$), above which LFEs were included in the finalized catalog and used for our analysis, should exceed the upper noise limit. Histograms of *CC*-values in Fig. 3b show an asymmetric distribution with a tall peak at *CC*~0.15. We followed previous studies (Green and Neuberg, 2006; Petersen, 2007; Lamb et al., 2015), in which the distribution of lower *CC*-values was modeled by a normally distributed curve that would be expected for random correlations between signals and noise, while the upper tail was considered to represent the presence of well-correlated LFEs. Visual inspection shows that frequencies at and below *CC*~0.15 are in good agreement with the left-hand side of the normally distributed curve where the mean is 0.155 and its standard deviation is 0.02 (Fig. 3b). We selected $CC_{th}$=0.25, which is larger than the mean plus three standard deviations. The histogram for $M \geq 0$ is also displayed because our analysis basically did not include LFEs with $M<0$. A total of 35 template LFEs were used for the finalized catalog.

The number of LFEs (894) in our finalized catalog of $CC_{th}$=0.25 (Fig. 3c) is about 19 times more than in the JMA catalog, which lists 47 LFEs that were detected in 2005-2020 by a conventional method that is not based on *CC*. Supplementary Fig. 4 shows an example of waveforms of a LFE: continuous waveforms and their matched template waveforms at each channel near the arrival times of the detected event. The *CC*-value in this figure is 0.39, nearly equal to the *CC*-value (*CC*=0.32) for an illustrative example showing continuous waveforms and template waveforms of a detected LFE in Shelly et al. (2007). Note that we did not detect ordinary earthquakes by ourselves, but merely used ordinary earthquakes in the JMA catalog, i.e., we used ordinary earthquakes detected by JMA. Given that analyzing waveforms of ordinary earthquakes was beyond the scope of our study, we felt that it was sufficient for this study to show an example of LFE waveforms.

The reader may think that $CC_{th}$=0.25 is a low cross-correlation value for the minimum threshold. However, assuming this value is not an issue, as is discussed next. The MF procedure used in the present study and our previous studies (Yukutake, 2017; Yukutake et al., 2019) is fundamentally the same as that used by Shelly et al. (2007), who detected LFEs in Shikoku, Japan. Our criterion of nine times the median absolute deviation (see section 2.2) is the same as that adopted by Shelly et al. (2007). This criterion corresponds to an exceedance probability of ~$6.4 \times 10^{-10}$ (Shelly et al., 2007). Fig. 3 of Shelly et al. (2007) demonstrated an illustrative example of a detected LFE with *CC*=0.32, showing continuous waveforms and template event waveforms for 30 traces (three components of 10 stations), where individual *CC*s varied from -0.14 to 0.69, and there was no high value ($\geq$0.85) for *CC* and individual *CC*s. Based on the previous study (Shelly et al., 2007), one might not consider that standard accepted *CC*-values are 0.85 and above. Shelly et al. (2007) stated that although individual *CC*s are modest, they are overwhelmingly positive and extremely unlikely to have occurred by chance. We can thus use *CC*-values below 0.5 and it is not necessary to use only *CC*-values greater than 0.85. Thus, there was no need to eliminate the LFE analysis based on the MF method from our study.

The scope of this study did not permit us to reveal repeating LFEs, nor cyclic activities and cluster characteristics, as were studied by Lamb et al. (2015). Rather, the aim of this study was to resolve the difficulty in detecting smaller LFEs. Our future research will conduct in-depth analyses of repeating LFEs for each cluster in the Izu-Tobu region, referring to Lamb et al. (2015), and using a sophisticated MF method that can locate detected LFEs to appreciate whether they occurred in the same cluster as the template LFE used to find them.

## 3. Methods

### 3.1. Change point analysis

The ETAS model (Ogata, 1988, 1992; Ogata and Tsuruoka, 2016) was originally introduced for ordinary earthquakes, but this model was also used for LFEs in the present study. This model treats cases in which an event $i$ that occurs at time $t_i$ triggers its offspring events; hence, the occurrence rate at time $t$ is given by the linear superposition of the clustering effects in the past. The sum of all events that occurred before $t$ is assumed. Seismicity cannot always be explained only by the clustering effects. In the framework of the ETAS model, such activity is called background activity and it is distinguished from clustering activity. Thus, the parameter set $\theta$ used in the ETAS model consists of five elements ($\mu$, $K_0$, $c$, $\alpha$, $p$). Here, $\mu$ is a parameter that characterizes the background activity, represented by the occurrence rate of this activity, where the unit is day$^{-1}$. $K_0$, $c$, $\alpha$, and $p$ are parameters that characterize clustering activity represented by the aftershock Omori–Utsu formula (Utsu 1961), where the units are day$^{-1}$, no unit, day, and no unit, respectively. In the standard ETAS model, $\theta$ is a time-independent constant during a study period.

We conducted a change point analysis using the ETAS model (Ogata and Tsuruoka, 2016) and the Akaike Information Criterion (AIC) (Akaike, 1974). Details of this analysis were provided by Kumazawa et al. (2010, 2019). The parameter from Kumazawa et al. (2010, 2019), which we used in this study, was $\Delta AIC=AIC_{single}-AIC_{2stage}$. Here, $AIC_{single}$ is the AIC for the standard (single) ETAS model that considers constant parameter values over time, and $AIC_{2stage}$ is the AIC for the two-stage ETAS model that considers different parameter values in subperiods before and after the change-point time $T_c$, where both models are fitted to LFEs of $M \geq M_{th}$ (threshold magnitude). If $\Delta AIC \geq 2q$ when plotting $\Delta AIC$ as a function of $T_c$, the two-stage model is better fitted to the data than the single model, detecting candidate changes in occurrence rate, where $q$ is the degree of freedom to search for the best candidate $T_c$ from the data. $q$ depends on sample size (number of LEFs) (Ogata, 1992; Kumazawa et al., 2010), and increases with sample size. Given that $\Delta AIC \geq 2q$ is observed, a change point's confidence interval of 68% was calculated (Kumazawa et al., 2010). To calculate it, we considered the likelihood of the two-stage ETAS model, $\exp\{-(AIC_{2stage}+2q)/2\}$ for different $T_0$-values. The normalization of the likelihoods for all $T_0$-values assigned a probability mass, and the central 68% of this mass provided a change point's confidence interval (error bounds). The same analysis was conducted for the time-dependent activity of ordinary earthquakes.

### 3.2. Choice of $M_{th}$

Analyses of the ETAS model of LFEs during a specified time interval are critically dependent on the choice of the $M_{th}$ value. To select it, we referred to estimates of completeness magnitude ($M_c$) of the processed data of LFEs. Above $M_c$, all LFEs are considered to have been detected. If $M_{th}$ is too low, compared to estimates of $M_c$, then it leads to an unreliable ETAS fitting.

To compute $M_c$, we employed the Entire-Magnitude-Range (EMR) technique (Woessner and Wiemer, 2005), based on the Gutenberg-Richter (GR) relation (Gutenberg and Richter, 1944), given by $\log_{10}N=a-bM$, where $N$ is the number of LFEs with a magnitude larger than or equal to $M$ in a given time window, and $a$ and $b$ are constants. Typically, $b$ is ~1 for ordinary earthquakes (e.g., Nanjo and Yoshida, 2018; Nanjo, 2020) and ~1.5 for LFEs (Nanjo et al., 2023). Statistical modeling distinguishes between completely detected and incompletely detected parts of the frequency-magnitude distribution (Woessner and Wiemer, 2005). $M_c$ is the magnitude at which these parts are separated. The $b$- and $a$-values are based on earthquakes in the completely detected part above $M_c$. It is important to address the uncertainty introduced by using a limited dataset and the GR-based method (Woessner and Wiemer, 2005). We computed uncertainty (one standard deviation) for $b$ and $M_c$ based on a bootstrapping technique (Schorlemmer et al., 2003).

The parameters obtained for LFEs in the entire catalog (2005-2020) were $b=1.47\pm0.09$, $a=2.79$, and $M_c=0.23\pm0.07$. The GR relation with $b=1.47$ and $a=2.79$ for $M\geq0.23$ is provided in Fig. 4a, showing a fit of the GR relation to observations in the present case. We selected $M_{th}=0.3$ ($M\geq0.3$) for analyses of the ETAS model of LFEs. We also considered $M_{th}=0.4$ ($M\geq0.4$) to suggest a generally stable feature.

Similar to the LFEs, we estimated $M_c$ values for ordinary earthquakes (Fig. 4b). The difference to the $M_c$ calculation for LFEs is that a single value of $M_c$ over the entire catalog was not considered. Instead, a timeseries of $M_c$, obtained from a moving window approach, was considered, where the window covered 300 ordinary earthquakes (horizontal light blue segment for each window). The difference in the $M_c$ calculation between ordinary earthquakes and LFEs is that this moving window approach was adopted for the former earthquakes and to calculate the local properties of an input data stream and an output variation in $M_c$. Smaller ordinary earthquakes during the period of a swarm are, in many cases, missing from earthquake catalogs as they are masked by larger earthquakes' coda and overlap on seismograms. In such a circumstance, the variation in $M_c$ may be large. Fig. 4b shows that $M_c$ values were generally about 0.8, except for the timing of the 2006 and 2009 swarms where $M_c$ values were greater than 1. To create this time-$M_c$ graph, uncertainty in $M_c$, indicated by a light blue vertical segment for each $M_c$ value (each time window), was computed based on a bootstrapping technique (Schorlemmer et al., 2003). We selected $M_{th}$=1 ($M \geq 1$) for analyses of the ETAS model of ordinary earthquakes. We also considered another value ($M_{th}$=1.5) for the same reason as that described for LFEs.

## 4. Results

### 4.1. First-order timeseries analysis

We conducted a first-order timeseries analysis based on the entire period from Jan. 2005 to Dec. 2020 (Fig. 5). For LFEs of $M \geq 0.3$ (blue data points in Fig. 5b), we observed the values of $\Delta AIC \geq 2q$, suggesting that the two-stage ETAS model is more applicable to $M \geq 0.3$ LFEs than the standard (single) ETAS model, where each blue horizontal dashed line indicates $2q$. The change point's confidence interval, in which the time to separate into the two-stages confidently fell, as indicated by a blue semitransparent region, was $T_c$=2725-3085 days. To define the change point's confidence interval, we first normalized the likelihoods every 5 days to assign a probability mass, and then considered the central 68% of this probability mass. The interval of the central 68% was defined as the change point's confidence interval (see section 3.1). $T_c$ with the largest $\Delta AIC$ (green circle in Fig. 5b) was 2740 days (Jul. 4, 2012). This fell within the change point's confidence interval, indicating that $T_c$=2740 days was one of the most probable change-point times. Note that the change point's confidence interval is not expected for periods of increased seismicity (swarms), but only for times when seismicity rate changed. Similarly, we observed that the change point's confidence interval for $M \geq 0.4$ was $T_c$=2545-3075 days (red semitransparent region). $T_c$ with the largest $\Delta AIC$ was 2560 days (Jan. 5, 2012).

The same analysis was conducted for ordinary earthquakes (Fig. 5a). The change point's confidence interval was $T_c$=1815-2425 days (blue semitransparent region) for $M \geq 1$ and $T_c$=1815-2345 days (red semitransparent region) for $M \geq 1.5$. Fig. 5a does not indicate that the change in occurrence rate occurred in both 2009 and 2011. The data informs us that the change in occurrence rate most likely occurred during the interval referred to as the change point's confidence interval (semitransparent region), even if readers might observe different values of ΔAIC following 2009 and 2011, as well as negative slopes of the relation between ΔAIC and time. $T_c$ with the largest ΔAIC (green circle) was 1820 days, corresponding to Dec. 26, 2009 for $M≥1$. $T_c$ with the largest ΔAIC for $M≥1.5$, was the same as that for $M≥1$. $T_c$=1820 days (Dec. 26, 2009), with the largest ΔAIC (green circle), fell within the change point's confidence interval. This result was one of the most probable change-point times and coincided with the timing of the 2009 swarm (blue vertical line). For $M≥1$ (blue data) and $M≥1.5$ (red data), we observed two other pronounced peaks in ΔAIC at the timing of the 2006 swarm (vertical orange line) and during 2011: explicitly, at $T_c$=480 days (ΔAIC=45.2 for $M≥1$ and 28.2 for $M≥1.5$) and $T_c$=2390 days (ΔAIC=75.5 for $M≥1$ and 33.7 for $M≥1.5$). A comparison between Fig. 5a and b shows that the change point's confidence intervals for ordinary earthquakes (Fig. 5a) were earlier than those for LFEs (Fig. 5b).

Fig. 5b shows better fitting of the two-stage ETAS model with a $T_c$-value falling in the change point's confidence interval to LFEs than the standard (single) ETAS model. This indicates that a change in occurrence rate occurred and that the change-point time $T_c$ fell sometime within this interval. The next question is which happened, relative activation or quiescence? To address this question, we used $T_c$=2740 days (green circle in Fig. 5b), corresponding to Jul. 4, 2012, as an illustrative example (Fig. 6c) and fitted the standard (single) ETAS model to LFEs ($M≥0.3$) before Jul. 4, 2012. The significance of $T_c$=2740 days (Jul. 4, 2012) is that this date was one of the most probable change point times. The occurrence rate (black) after Jul. 4, 2012 was smaller than the extrapolated rate (red), showing relative quiescence, where the extrapolated rate is the occurrence rate computed by using the standard (single) ETAS model whose θ was obtained by fitting this model to LFEs before Jul. 4, 2012. To confirm this quiescence, LFEs after Jul. 4, 2012 were fitted by the standard (single) ETAS model (Fig. 6d). The occurrence rate (black) before Jul. 4, 2012 was smaller than the extrapolated rate (red), where the extrapolated rate was computed by using the standard (single) ETAS model with θ obtained by fitting of this model to LFEs after Jul. 4, 2012.

The same analysis was conducted for ordinary earthquakes. Similar to LFEs, we used $T_c$=1820 days (green circle in Fig. 5a), corresponding to Dec. 26, 2009, as a representative example (Fig. 6a). We then fitted the standard (single) ETAS model to ordinary earthquakes ($M≥1$) before Dec. 26, 2009. Fig. 6a shows that the occurrence rate (black) after Dec. 26, 2009 was smaller than the extrapolated rate (red), indicating relative quiescence. This quiescence (Fig. 6b) was confirmed by fitting the standard (single) ETAS model to ordinary earthquakes after Dec. 26, 2009. Fig. 6b shows that the occurrence rate (black) before Dec. 26, 2009 was smaller than the extrapolated rate (red).

The above results were not induced by including the period around the timing of the first swarm in 2006 into the study period (Jan. 2005-Dec. 2020). Namely, we conducted a timeseries analysis of ordinary earthquakes and LFEs during the subperiod Jan. 2008-Dec. 2020. Results show that the general ΔAIC-$T_c$ patterns (Fig. 7) for both types of earthquakes were similar to those during the corresponding period in Fig. 5. The change point's confidence interval coincided with the timing of the 2009 swarm for ordinary earthquakes (Fig. 7a), while for LFEs, it emerged during the period from mid-2012 to mid-2013 for LFEs (Fig. 7b).

*4.2. Second-order timeseries analysis*

We conducted a second-order timeseries analysis, based on the subperiods Jan. 2005-Dec. 2010 for LFEs (Fig. 8b) and Jan. 2005-Dec. 2008 for ordinary earthquakes (Fig. 8a). We considered these subperiods because they were before the change point's confidence intervals seen in the results for the entire study period (Fig. 5). The results of ordinary earthquakes (Fig. 8a) show that the change point's confidence interval ($T_c$=475-485 days for $M_{th}$=1.0 and 1.5) was limited to the timing of the 2006 swarm (vertical orange line).

The results of LFEs with $M_{th}$=0.3 (Fig. 8b) show the change point's confidence interval ($T_c$=585-795 days) after the 2006 swarm. ΔAIC for LFEs with $M_{th}$=0.4 (red data in Fig. 8b) shows a better but insignificant outcome when the two-stage ETAS model, rather than the single ETAS model, was used. Namely, ΔAIC was higher than 0 (ΔAIC>0), but it was below the horizontal dashed line, which is a hurdle to the selection of the two-stage ETAS model. We interpreted this insignificant outcome as an indication that the number of LFEs with $M≥0.4$ was not enough to achieve the desired conclusion. We do not show the change point's confidence interval for $M≥0.4$.

A detailed observation shows that the confidence interval was later for LFEs than for ordinary earthquakes. Fig. 9 shows that the change in occurrence rate at $T_c$ (green circle in Fig. 8) for the two types of earthquakes was relative quiescence.

Overall, the first-order observation was that quiescence for both types of earthquakes started during or after the 2009 swarm. A detailed observation showed that the start of quiescence was earlier for ordinary earthquakes than for LFEs. Beyond that, we showed that second-order quiescence occurred before the start of first-order quiescence. This second-order quiescence started during or after the 2006 swarm. Similar to the first-order timeseries analysis, a detailed observation showed an earlier start of quiescence for ordinary earthquakes than for LFEs.

*4.3. Background rate*

In the framework of the ETAS model, all earthquakes are decomposed into background activity and clustering activity. Kumazawa et al. (2016), who measured the occurrence rate of the former activity (µ) separated by that of the latter activity, studied eight swarms of ordinary earthquakes after mid-1980 to associate µ with volumetric strain recorded by the strainmeter located at the Higashi-Izu JMA station (Fig. 1b). Given that the pattern of strain variation at this station depends on magma intrusion (ERC, 2010), Kumazawa et al. (2016) proposed that the ETAS model and µ may help in monitoring magma intrusions that drive changes in stress. Motivated by previous studies (ERC, 2010; Kumazawa et al., 2016), we examined whether or not µ changed with time.

This study adopted a simple approach to capture essential aspects of the time-dependent occurrence rate of background activity (µ), although a sophisticated nonstationary model is available (e.g., Kumazawa et al., 2016, 2019). We adopted a time-window approach, considering the results of LFEs obtained from the first-order timeseries analysis (Jan. 2005-Dec. 2020), and comparing µ between the periods before and after $T_c$=2740 days (green circle in Fig. 5b). The results of LFEs (Fig. 6b) showed a larger µ-value for the period before $T_c$=2740 days (µ=0.046) (Fig. 6b) than for the period after it (µ=0.022). Similarly, the results of ordinary earthquakes (Fig. 6a) showed that µ=0.019 before $T_c$=1820 days (green circle in Fig. 5a) was larger than µ=0.003 after it. The same comparison in µ was conducted for the second-order timeseries analysis of LFEs (Jan. 2005-Dec. 2010) and ordinary earthquakes (Jan. 2005-Dec. 2008). The result, i.e., µ=0.069 before $T_c$=620 days (green circle in Fig. 8b) and µ=0.029 after it for LFEs (Fig. 9b) and µ=0.024 before $T_c$=480 days (green circle in Fig. 8a) and µ=0.010 after it for ordinary earthquakes (Fig. 9a), shows a similar feature to that for the first-order timeseries analysis. We associated the seismic quiescence observed in the first- and second-order timeseries analysis with decreases in µ.

5. Discussion

To compare the characteristics of the two types of earthquakes in the study region, we discuss crustal movement data. A leveling survey along the eastern coast of the Izu Peninsula since 1904 until 2015 (Fig. 10) showed secular variation of four benchmarks (9335, 9336, 9337, 9338), where the reference benchmark is 9328 (GSI, 2016). The former four benchmarks were located in the Izu-Tobu region while the latter reference benchmark was located outside of the northern periphery of this region. The upward and downward crustal movement of the four benchmarks, relative to the reference benchmark, is interpreted as an indication of uplift and subsidence of the Izu-Tobu region, respectively. According to the GSJ (2016), the moments of seismic events were indicated by arrows and numbered by 1, 2, … 14 (Fig. 10a). Also included in Fig. 10a is the time period (grey region) that was currently analyzed. Fig. 10b shows where the numbered seismic events occurred in order to include these seismic events alongside the leveling data (Fig. 10a). A brief summary of crustal movement in the Izu-Tobu region is provided next. The uplift progressed during the 1920s-1930s, where the 1923 *M*7.9 Kanto earthquake (1) and the 1930 *M*7.3 Kita Izu earthquake (3) occurred outside the Izu-Tobu region, while the 1930 swarm (2) occurred inside that region. Until the early 1970s, pronounced crustal movement was not observed. Around the timing of the 1974 Izu Hanto-Oki earthquake (8), the uplift started to make noticeable progress and continued until the late 1990s, after which the uplift abated. After the 2009 swarm (14) ended, an uplift was not observed.

Graphs of relative height of the four stations, numbered 1, 2, 3, and 4 in the reference stations ODAWARA (Fig. 11a) and HATSUSHIMA (Fig. 11b) show no uplift, even after 2015, which is the last year of the leveling survey shown in Fig. 10. We selected the four stations and the two reference stations to consider the baselines, because we attempted to mimic the configuration of the benchmarks used for the leveling survey (inset of Fig. 10). To create Fig. 11, we used station coordinates derived from the GNSS Earth Observation Network System (GEONET) (Muramatsu et al., 2021; Takamatsu et al., 2023). The product in the current GEONET analysis strategy is called the F5 solution while that in the previous strategy is called the F3 solution. Blue and red data in Fig. 11 were based on the F5 and F3 solutions, respectively. The F5 solution was used for the period Oct. 2013-Dec. 2020 because data in the most recent 10-year period were obtained. To complement data since Jan. 2005 (start of our study period), the F3 solution was used for the period Jan. 2005-May 2016. The F3 and F5 solutions overlapped during the period Oct. 2013-May 2016, providing a justification for the consistency between these solutions (Fig. 11). Similarly, Fig. 12 was created for the reference station YUGAWARA-A, which was in operation since the end of Mar. 2017.

For all of the baselines, fluctuations in relative height were observed, and some of them were associated with the replacement of antennas at stations (triangle) and tree trimming around stations (arrow). We also observed an increase in the order of a few centimeters in relative height around the timing of the 2006 swarm in panel 4 of Fig. 11 (baselines ITOUYAHATANO-ODAWARA and ITOUYAHATANO-HATSUSHIMA), as seen in the leveling survey for benchmark 9338 (red curve in Fig. 10). However, the tendency for relative heights to remain unchanged over time (no increasing nor decreasing trend in relative height) for all of the baselines was generally observed. Combining the results of the leveling survey (Fig. 10) and those of baseline changes (Figs. 11 and 12) shows no significant uplift in the study period (Jan. 2005-Dec. 2020).

It is not surprising that no significant uplift and subsidence were observed in the study region. A study based on morphology, stratigraphy, and fossil assemblages in the eastern side of the Izu

Peninsula (Shishikura et al., 2023) identified three events over 1500 years: a 1.05-m uplift in 595-715, a 1.33-m uplift in 1356-1666, and a 0.82-m uplift after 1830. The interval between events was 400-800 years. A cumulative uplift of 0.6-0.9 m since 1904, shown in Fig. 10, plus an uplift of 0.1-0.2 m in 1868 or 1870 (Koyama, 1999) before the start of instrument-based measurements, roughly coincided with the uplift of 0.82 m after 1830 (Shishikura et al., 2023). This last uplift was mainly caused by vertical movement due to subsurface magmatic movement accompanied by earthquake swarms, while the first and second uplifts were considered to be caused by similar volcanic deformation or coseismic deformation due to offshore faults. There were two periods (715-1356 and 1666-1830) of no uplift during 1500 years on the eastern side of the Izu Peninsula.

A comparison between crustal movement data (Figs. 10-12) and seismicity data (Figs. 5-9) showed that there was no significant uplift during the study period (2005-2020), including time periods of the two swarms. Our observation is not surprising for an active volcanic area. LFEs beneath the active volcano Mt. Fuji activated after the $M$5.9 Shizuoka earthquake that occurred on Mar. 15, 2011 at the foot of the volcano, and the occurrence rate of LFEs did not return to pre-earthquake levels (Nanjo et al., 2023). No crustal deformation during the $M$5.9 event was reported.

The first-order observation based on the entire period showed that the change point's confidence intervals for both types of earthquakes were during or after the 2009 swarm, indicating that quiescence started during or after the 2009 swarm (Figs. 5-7). Similarly, the second-order observation based on the subperiods before the change point's confidence intervals indicated by the first-order observation showed that quiescence for both types of earthquakes started during and after the 2006 swarm (Figs. 8 and 9). These results imply that the start of quiescence for both types of earthquakes was associated with the 2006 and 2009 swarms.

We further pointed out the relation between quiescence and a decrease in μ (section 4.3) for both types of earthquakes, suggesting that a decrease in background activity might cause quiescence. Beyond the standard (single) ETAS model, μ means the rate of earthquake occurrence caused by sources that can vary with time (Kumazawa et al., 2016). Past studies (Hainzl and Ogata, 2005; Llenos et al., 2009; Llenos and McGuire, 2011; Brodsky and Lajoie, 2013) indicated that variation in stress affects μ. Based on these previous studies, the observed decrease in μ (section 4.3) allows us to assume that there exist sources that can vary with time, resulting in changes in stress.

Both first- and second-order timeseries cases revealed that the start of quiescence for ordinary earthquakes was earlier than for LFEs (Figs. 5 and 9). This implies that the changes in occurrence rate of ordinary earthquakes in a shallow depth range were associated with those of LFEs in a deep depth range. Our study showed no evidence that the 2006 and 2009 swarms were definitely magmatic swarms because we did not obtain images of consistent intrusion into the study region using tomography or other geophysical methods. However, previous studies summarized in the Introduction showed some evidence of past swarms (Ishida, 1984; Shimazaki, 1988; Tada and Hashimoto, 1989; Oshima et al., 1991; Yamamoto et al., 1991; Nagamune et al., 1992; Okada and Yamamoto, 1991; Okada et al., 2000; Hayashi and Morita, 2003; Morita et al., 2006; Miyamura et al., 2010; Ukawa, 1991; Miyamura et al., 2010; ERC, 2010; GSI, 2016; Kumazawa et al., 2016; Shishikura et al., 2023). If these previous studies are considered valid, then we propose that the changes in occurrence rate of shallow ordinary earthquakes and those of deep LFEs were caused by magma.

Typically, one would expect magma to move from a depth toward a crustal magma reservoir, first causing a change in occurrence rate of deep LFEs. However, our study showed that there was a lag between the early timing of change in occurrence rate (quiescence) for shallow ordinary earthquakes and a late timing for deep LFEs (Figs. 5-9). We provide an interpretative hypothesis next.

First, magma has not been supplied from a depth toward a crustal magma reservoir since the 2000s, causing no significant uplift since the 2000s (Fig. 10). Second, in the absence of a magma supply from a depth, magma intruded into the shallow crust from the crustal magma reservoir, and stresses caused by the magma intrusion into the shallow crust would explain swarms such as the 2006 and 2009 swarms that occurred in the shallow crust (Fig. 2). Third, due to the intrusion of magma from this reservoir into the shallow crust, a decrease in pressure of magma fluid in shallow parts of the reservoir resulted in a decrease in stress in and around these parts. Fourth, because of this decrease in stress, quiescence (decrease in $\mu$) for shallow ordinary earthquakes started during or after each of the two swarms (Figs. 5a and 8a). Finally, the decrease in pressure of magma fluid propagated from shallow to deep parts of the crustal magma reservoir caused a decrease in stress in and around the deep parts. Thus, there existed a lag between an early start of quiescence for shallow ordinary earthquake (Figs. 5a and 8a) and a late start for deep LFEs (Figs. 5b and 8b).

Earthquake swarms have been attributed to stress perturbations by magma intrusions (Einarsson and Brandsdottir, 1980; Dieterich et al., 2000; Okada et al., 2000; Toda et al., 2002; Waite and Smith, 2002; Feuillet et al., 2004; Smith et al., 2004) or fluid injections (Hainzl and Ogata, 2005; Lei et al., 2008; Terakawa et al., 2013; Terakawa, 2014). On the other hand, they can also be triggered by creep or slow-slip events (Ozawa et al., 2003; Delahaye et al., 2009; Segall et al., 2006). Thus, swarms are a direct response to the changes in stress caused by intrusion and tectonic events. Swarms caused by changes in fluid pressure require a stress source to change pressure, and earthquakes are a seismic response to that change in fluid pressure.

Swarms also occur in environments unrelated to magmatism (Harath et al., 2022). Swarms occur when lakes form behind a newly constructed dam or the removal of a dam. Swarms are frequently observed at the initial impoundment stage and manifest as elevated seismic rates in the shallow crust, which decays rapidly (within months to a few years), relative to pre-impoundment background rates. This establishes a straightforward causal relationship between crustal poroelastic response and perturbations to an ambient stress field from instantaneous loading and pore pressure diffusion (Simpson, 1976). In several other instances, delayed pulsating episodes of reservoir-triggered seismicity associated with crustal response to periodic fluctuations in reservoir water level have been identified (Simpson et al., 1988; Telesca, 2010). These examples indicate a limit of our discussion to the specific topic under study in this paper.

## 6. Conclusions

This paper examined earthquake data in the Izu-Tobu region located within the Izu Peninsula, Japan during Jan. 2005-Dec. 2020. The primary purpose of this study was to better understand the characteristics of the time-dependent activity of LFEs in the region during the same period. We applied a *CC*-based method (MF method) to identify LFEs within the seismic activity. This paper provides a detailed description of the measures adopted to optimize the method and minimize the risk of misdetection of LFEs, particularly due to the influence of teleseismic events. The number of LFEs (894) in our finalized catalog (Fig. 3c) was about 19 times larger than that of LFEs (47) in the JMA catalog, which was created by using a conventional method that is not based on *CC*. We applied ETAS modeling to both deep LFEs and shallow ordinary (high frequency) earthquakes. The output obtained from this study showed relative quiescence of seismic activity (Figs. 5-9).

The first-order timeseries analysis of LFEs during the study period (Jan. 2005-Dec. 2020) was conducted using the standard (single) ETAS model and an alternative two-stage ETAS model, considering different parameter values in subperiods before and after the change point time $T_c$. The latter two-stage model was significantly better than the former standard (single) model, when $T_c$ fell in the period of early 2012 to mid-2013 (semitransparent regions in Fig. 5b).

We were interested to assess whether there was a relationship between LFEs and ordinary earthquakes. For comparison, the same analysis was conducted for ordinary earthquakes, showing a similar result with $T_c$ falling in the period of late 2009 to mid-2011 (semitransparent regions in Fig. 5a). Our results showed an earlier $T_c$ for ordinary earthquakes than for LFEs (Fig. 5). Further analyses using the standard (single) ETAS model revealed that quiescence occurred for both types of earthquakes (Fig. 7).

Beyond those findings, we observed that second-order quiescence occurred before the start of first-order quiescence (Figs. 8 and 9). Second-order quiescence started around the timing of the 2006 swarm for ordinary earthquakes and after the 2006 swarm for LFEs (Fig. 8).

Motivated by previous studies (ERC, 2010; Kumazawa et al., 2016), we showed that background activity characterized by μ changed with time for both types of earthquakes (section 4.3). Moreover, we proposed that quiescence was associated with a decrease in μ.

Our results were compared with crustal movement data (Figs. 10-12) to show that seismic quiescence for both types of earthquakes occurred without significant uplift and subsidence in and around the study region (Izu-Tobu region) during the study period (2005-2020). For this comparison, surface vertical displacement data obtained from a leveling survey and GEONET were used (Figs. 10-12). The uplift was significant during the 1970s-1990s while it was in abatement or unobservable in the late 1990 and later.

We propose an interpretative hypothesis regarding the relationship between the two types of earthquakes, as follows:

1. No supply of magma from a depth toward a crustal magma reservoir occurred since the 2000s, causing no significant uplift (Fig. 10).
2. Under the condition in 1, magma intrusion from this reservoir into the shallow crust and stress perturbation by the intrusion was attributed to the shallow 2006 and 2009 swarms (Fig. 2).
3. Due to the magma intrusion in 2, a decrease in pressure of magma fluid in shallow parts of the reservoir caused a decrease in stress in and around these parts.
4. Due to the decrease in stress in 3, quiescence (decrease in μ) for shallow ordinary earthquakes started during or after each of the two swarms (Figs. 5a and 8a).
5. The decrease in magma-fluid pressure in 3, which propagated into deep parts of the reservoir, induced a time lag between the early start of quiescence for shallow ordinary earthquakes and the late start for deep LFEs (Figs. 5b and 8b).

It is known that swarms also occur in environments unrelated to magmatism (Harath et al., 2022). Thus, our hypothesis provided above was limited to the specific topic that was studied in this paper.

**Data availability**

The datasets used and/or analyzed during the current study are available from the corresponding author upon reasonable request. The JMA catalog was obtained from https://www.data.jma.go.jp/eqev/data/bulletin/hypo.html. The waveform records were obtained from the permanent stations of the National Research Institute for Earth Science and Disaster Resilience, Earthquake Research Institute at the University of Tokyo, JMA, and the Hot Springs Research Institute of the Kanagawa Prefectural Government. The location of Teishi Knoll, used for Figs. 1 and 2 and Supplementary Figs. 1-3, was obtained from https://www.mri-jma.go.jp/Dep/sei/fhirose/plate/en.PlateData.html. The location of the Higashi-Izu JMA station, used for Fig. 1 and Supplementary Fig. 1, was obtained from Kumazawa (2016). The authors digitized the fault models for the 2006 swarm (GSI, 2007) and the 2009 swarm (JMA, 2010) to create Figs. 1 and 2 and Supplementary Figs. 1-3. The seismicity analysis software ZMAP (Wiemer, 2001), used for computing $M_c$ (Woessner and Wiemer, 2005) in Fig. 4, was obtained from https://github.com/swiss-seismological-service/zmap7. Data shown in Fig. 10 were reproduced from GSI (2016). Generic Mapping Tools (GMT) (Wessel et al., 2013), used for Figs. 1-3, 10-12, and Supplementary Figs. 1-3, is an open-source collection (https://www.generic-mapping-tools.org). GEONET data (Muramatsu et al., 2021; Takamatsu et al., 2023), used for Figs. 11 and 12, were obtained from https://mekira.gsi.go.jp/index.en.html.

**CRediT authorship contribution statement**

**K. Z. Nanjo**: Conceptualization, Investigation, Resources, Project administration, Formal analysis, Methodology, Software, Validation, Writing-original draft, Writing-review & editing. **Y. Yukutake**: Conceptualization, Investigation, Data curation, Resources, Formal analysis, Writing-review & editing. **T. Kumazawa**: Investigation, Methodology, Software, Writing-review & editing.

**Declaration of Competing Interests**

The authors declare that they have no known competing financial interests or personal relationships that could have appeared to influence the work reported in this paper.

**Acknowledgments**

This study was partially supported by the Ministry of Education, Culture, Sports, Science and Technology (MEXT) of Japan, under The Second Earthquake and Volcano Hazards Observation and Research Program (Earthquake and Volcano Hazard Reduction Research) (K.Z.N., Y.Y.) and under STAR-E (Seismology TowArd Research innovation with data of Earthquake) Program Grant Number JPJ010217 (K.Z.N., T.K.), JSPS KAKENHI Grant Numbers JP 22K03752 (Y.Y.), 20K11704 (T.K.), and a Research Grant of the Izu Peninsula UNESCO Global Geopark (K.Z.N., Y.Y.). The authors thank the Editor (S. Calvari) and the two anonymous reviewers for their constructive comments, and Y. Noda for help with implementing the MF method.

Lamb, O.D., De Angelis, S., Umakoshi, K., Hornby, A.J., Kendrick, J.E., Lavallée, Y., 2015. Repetitive fracturing during spine extrusion at Unzen volcano, Japan. Solid Earth 6, 1,277-1,293. https://doi.org/10.5194/se-6-1277-2015.

Lei, X., Yu, G., Ma, S., Wen, X., Wang, Q., 2008. Earthquakes induced by water injection at ∼3 km depth within the Rongchang gas field, Chongqing, China. Journal of Geophysical Research 113(B10), B10310. https://doi.org/10.1029/2008JB005604.

Llenos, A.L., McGuire, J.J., 2011. Detecting aseismic strain transients from seismicity data. Journal of Geophysical Research 116, B06305. https://doi.org/10.1029/2010JB007537.

Llenos, A.L., McGuire, J.J., Ogata, Y., 2009. Modeling seismic swarms triggered by aseismic transients. Earth and Planetary Science Letters 281(1-2), 59-69. https://doi.org/10.1016/j.epsl.2009.02.011.

Minakami, T., Ishikawa, T., Yagi, K., 1951. The 1944 Eruption of Volcano Usu in Hokkaido, Japan, History and mechanism of formation of the new dome "Showa-Sinzan". Bulletin of Volcanology 11, 45-157. https://doi.org/10.1007/BF02596029

Miyamura, J., Ueno, H., Yokota, T., 2010. Estimation of amount of intrusive magma by using volumetric strain data and attempt of evaluation of volcanic activity in Izu-Tobu volcanoes. Geophysical bulletin of Hokkaido University 73, 239-255. http://doi.org/10.14943/gbhu.73.239.

Moriwaki, K., 2017. Automatic detection of low-frequency earthquakes in southwest Japan using matched-filter technique. Journal of Seismology 81, 3 (in Japanese with English abstract). https://www.jma.go.jp/jma/kishou/books/kenshin/vol81_3.pdf.

Morita, Y., Nakao, S., Hayashi, Y., 2006. A quantitative approach to the dike intrusion process inferred from a joint analysis of geodetic and seismological data for the 1998 earthquake swarm off the east coast of Izu Peninsula. Journal of Geophysical Research 111, B06208. https://doi.org/10.1029/2005JB003860.

Muramatsu, H., Takamatsu, N., Abe, S, Furuya T., Kato C., Ohno, K., Hatanaka, Y., Kakiage, Y., Ohashi, K., 2021. Updating daily solution of CORS in Japan using new GEONET 5th analysis strategy. Journal of the Geospatial Information Authority of Japan 134, 19-32 (in Japanese). https://doi.org/10.57499/JOURNAL_134_03.

Nagamune, T., Yokoyama, H., Fukudome, A., 1992. Earthquake swarms off the east coast of the Izu peninsula and their relation to the 1989 eruption of Teishi Knoll volcano. Kazan (Bulletin of the

Ogata, Y., 1992. Detection of precursory relative quiescence before great earthquakes through a statistical model. Journal of Geophysical Research 97, 19845-19871. https://doi.org/10.1029/92JB00708.

Ogata, Y., Tsuruoka, H., 2016. Statistical monitoring of aftershock sequences: a case study of the 2015 Mw7.8 Gorkha, Nepal, earthquake. Earth, Planets and Space 68, 44. https://doi.org/10.1186/s40623-016-0410-8.

Okada, Y., Yamamoto, E., 1991. Dyke intrusion model for the 1989 seismovolcanic activity off Ito, Central Japan. Journal of Geophysical Research 96(B6), 10361-10376. https://doi.org/10.1029/91JB00427.

Okada, Y., Yamamoto, E., Ohkubo T., 2000. Coswarm and preswarm crustal deformation in the eastern Izu Peninsula, central Japan. Journal of Geophysical Research 105(B1), 681-692. https://doi.org/10.1029/1999JB900335.

Omori, F., 1911. The Usu-san eruption and earthquake and elevation phenomena. Bulletin of the Imperial Earthquake Investigation Committee 5, 1-38.

Oshima, S., Tsuchide, M., Kato, S. Okubo, S., Watanabe, K., Kudo, K. Ossaka, J., 1991. Birth of a submarine volcano "Teisi Knoll". Journal of Physics of the Earth 39(1), 1-19. https://doi.org/10.4294/jpe1952.39.1.

Ozawa, S., Miyazaki, S., Hatanaka, Y., Imakiire, T., Kaidzu, M., Murakami, M., 2003. Characteristic silent earthquakes in the eastern part of the Boso peninsula, Central Japan. Geophysical Research Letters 30(6), 1283. https://doi.org/10.1029/2002GL016665.

Peng, Z., Zhao, P., 2009. Migration of early aftershocks following the 2004 Parkfield earthquake. Nature Geoscience 2, 877-881. https://doi.org/10.1038/ngeo697.

Petersen, T., 2007. Swarms of repeating long-period earthquakes at Shishaldin Volcano, Alaska, 2001-2004. Journal of Volcanology and Geothermal Research 166(3-4), 177-192. https://doi.org/10.1016/j.jvolgeores.2007.07.014.

Power, J.A., Stihler, S.D., White, R.A., Moran, S.C., 2004. Observations of deep long-period (DLP) seismic events beneath Aleutian arc volcanoes; 1989-2002. Journal of Volcanology and Geothermal Research 138(3-4), 243-266. https://doi.org/10.1016/j.jvolgeores.2004.07.005.

Ross, Z.E., Idini, B., Jia, Z., Stephenson, O.L., Zhong, M., Wang, X., Zhan, Z., Simons, M., Fielding, E.J., Yun, S.-H., Hauksson, E., Moore, A.W., Liu, Z., Jung, J., 2019. Hierarchical interlocked
26

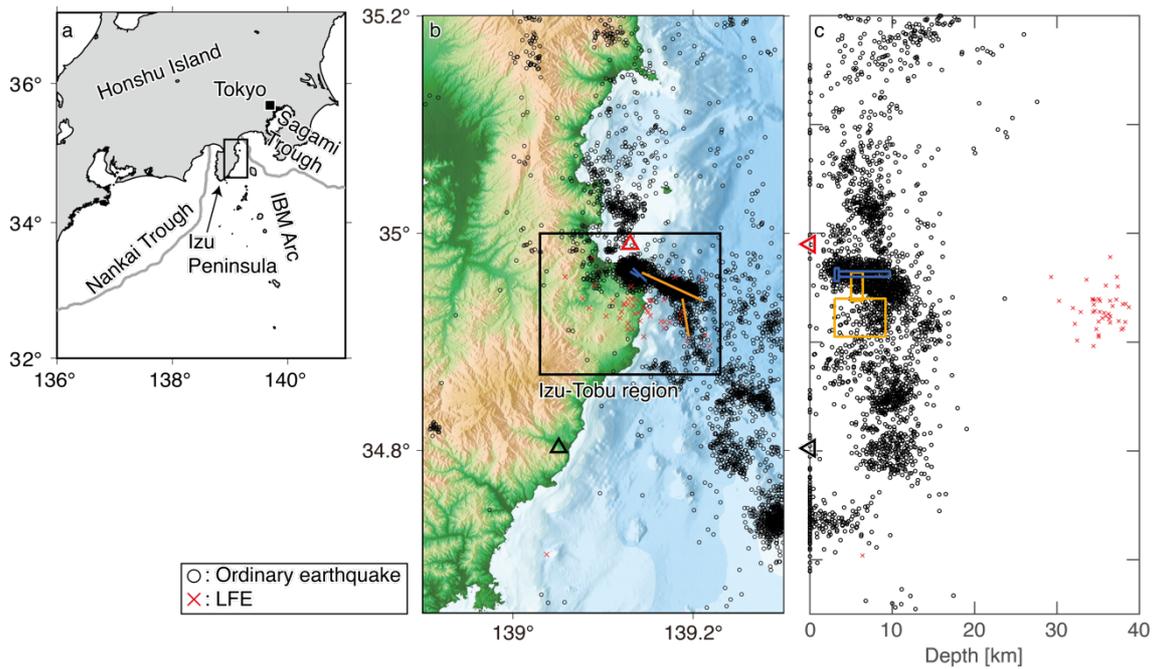

**Fig. 1.** Seismicity in and around the study region. **a**, Central Japan including the region (black rectangle) of **b**. Gray curves indicate trough axes. IBM Arc: Izu-Bonin-Mariana Arc. **b**, Map showing ordinary earthquakes with $M≥1$ (black circle) and LFEs with $M≥0.1$ (red cross) at depths 0-40 km during the period Jan. 2005-Dec. 2020. To plot these earthquakes, the JMA catalog was used. Orange and blue segments show the fault models for the 2006 and 2009 swarms, respectively (GSI, 2007; JMA, 2010). Red triangle indicates the Teishi Knoll. Black rectangle indicates the study region, known as the Izu-Tobu region. Black triangle indicates the Higashi-Izu JMA station, where a strainmeter that recorded volumetric strain data used by Kumazawa et al. (2016) was installed. **c**, Cross-sectional view of ordinary earthquakes and LFEs. Orange and blue rectangles show the fault models for the 2006 and 2009 swarms, respectively (GSI, 2007; JMA, 2010).

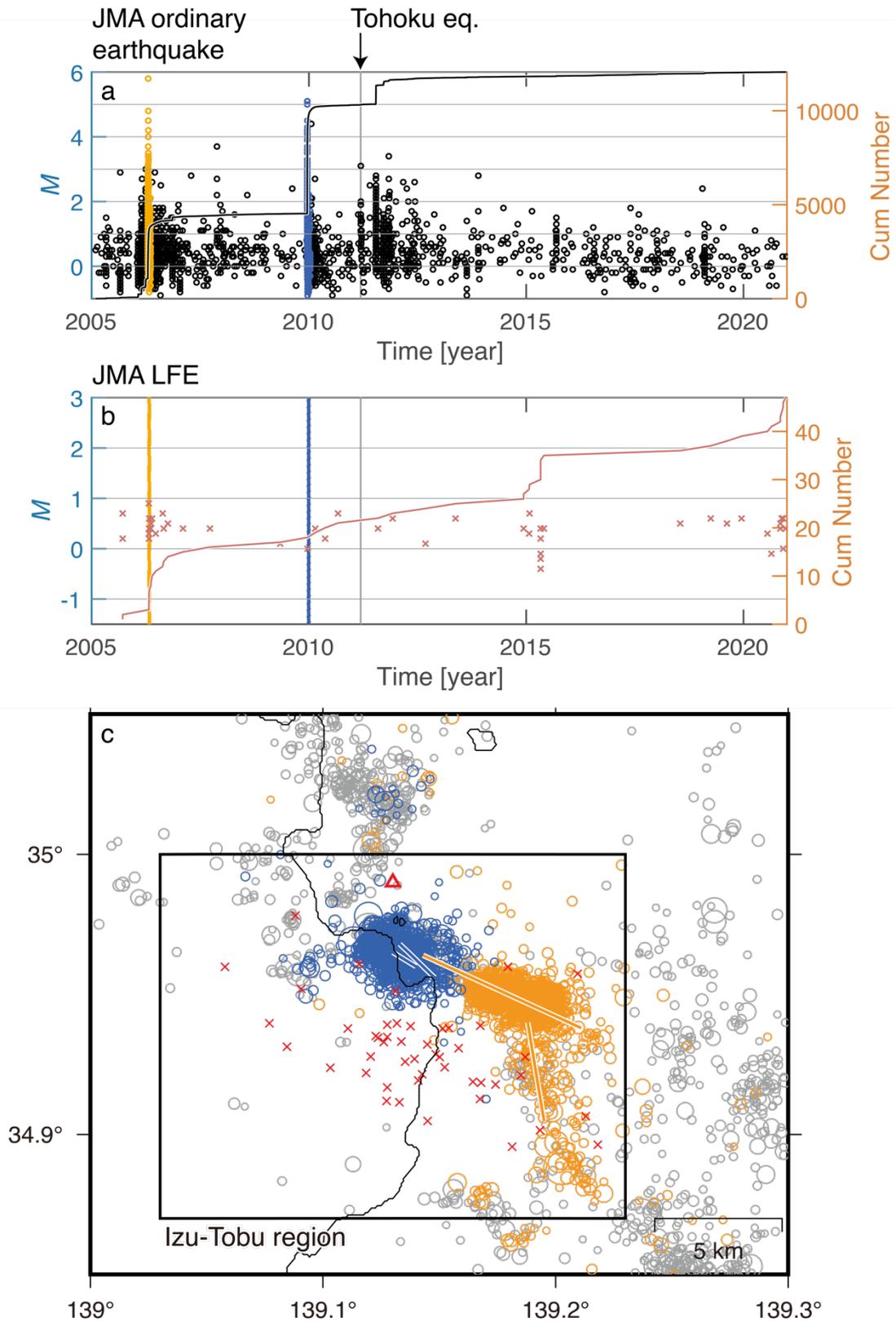

**Fig. 2.** Space-time distribution of earthquakes in the Izu-Tobu region, using the JMA catalog. **a**, *M*-time diagram (y-axis on the left side) of ordinary earthquakes (depths of 0-40 km) in the study

region indicated by a black rectangle in **c**. Overlapped is the cumulative number of ordinary earthquakes as a function of time (y-axis on the right side). Two abrupt increases in the cumulative number of ordinary earthquakes indicate two earthquake swarms. Based on JMA (2014), earthquakes in the periods Apr. 17-May. 12, 2006 (orange) and Dec. 16, 2009-Jan. 12, 2010 (blue) were considered as the 2006 and 2009 swarms, respectively. Vertical line indicates the moment of the Mar. 11, 2011 *M*9 Tohoku earthquake. **b**, Same as the top panel for LFEs. Orange and blue vertical lines indicate the 2006 and 2009 swarms, respectively. **c**, Map showing ordinary earthquakes (circles) and LFEs (crosses) in and around the study region (black rectangle). Earthquakes with $M \geq 1$ (depths of 0-40 km) in orange and blue are the 2006 and 2009 swarms, respectively. Other ordinary earthquakes are indicated by grey circles. Triangle indicates the most recent volcanic center, the Teishi Knoll. It is possible to observe the locations of the 2006 and 2009 swarms relative to the volcanic center.

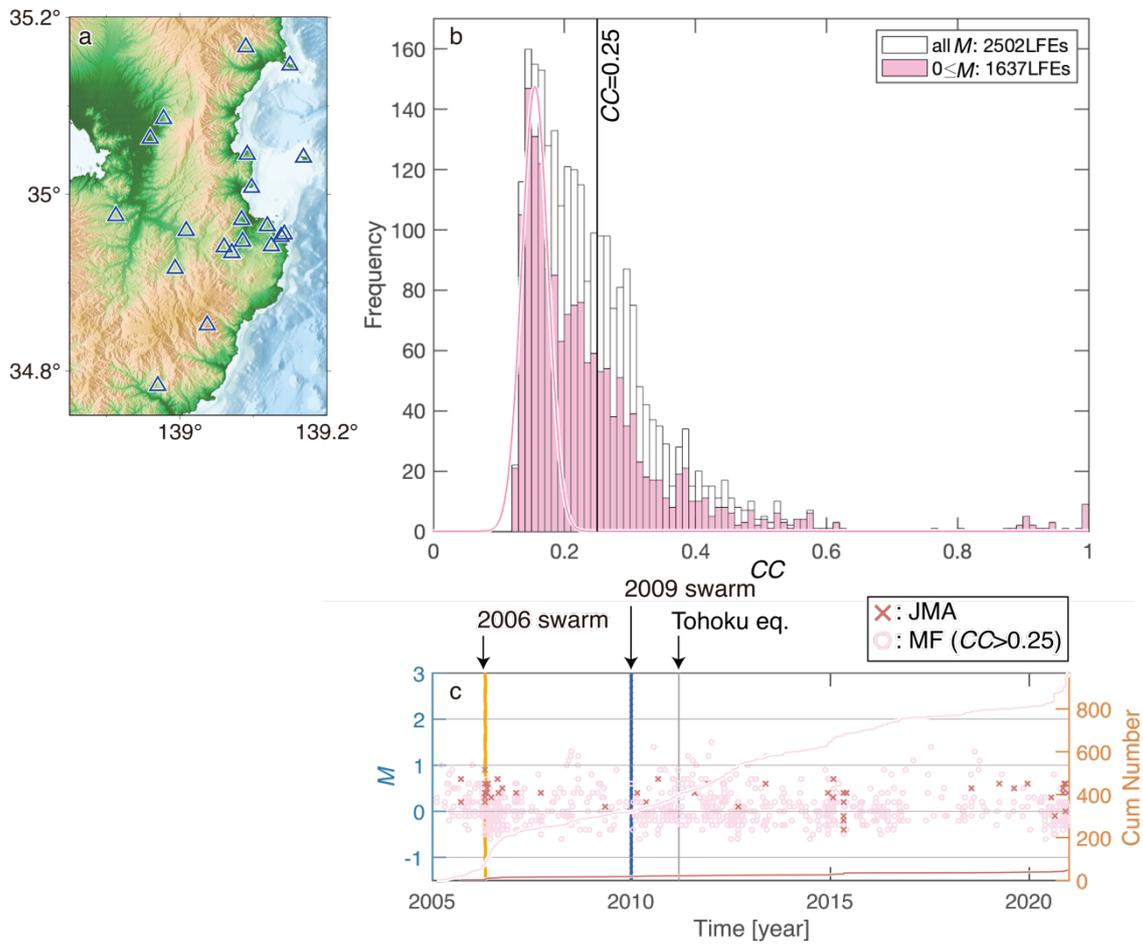

**Fig. 3.** LFEs detected by the MF method. **a**, Stations that recorded waveforms used in this study. **b**, Multiple histograms of *CC*-values for all magnitudes (white) and $M \geq 0$ (pink). Also included is the normally distributed curve (mean of 0.155 and standard deviation of 0.02). Vertical line indicates *CC*=0.25. **c**, Same as the bottom panel of Fig. 2**a**, but data from the MF catalog (*CC*>0.25) are included, as shown in pink.

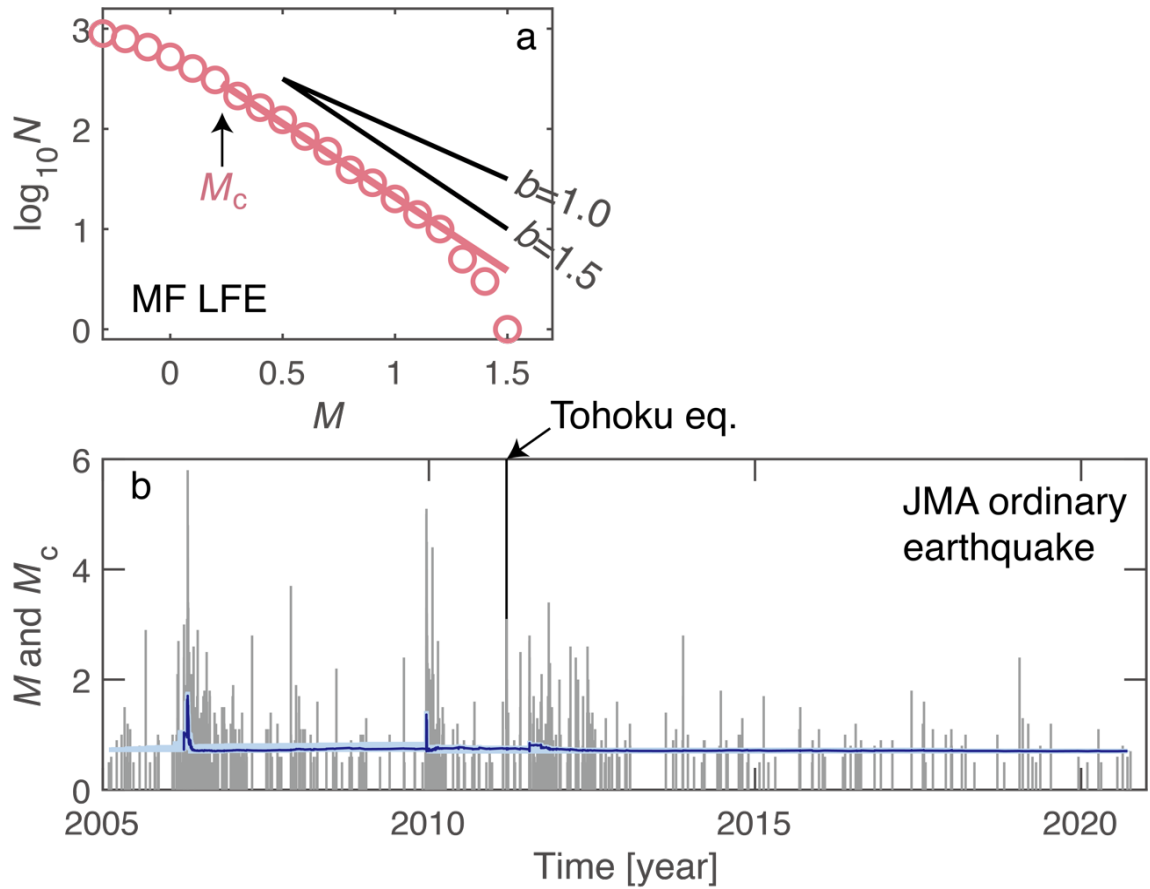

**Fig. 4.** $M_c$ of LFEs and ordinary earthquakes. **a,** The MF catalog ($CC>0.25$) for 2005-2020 was used. The GR relation with $b=1.47$ and $a=2.79$ for $M \geq M_c=0.23$ is shown. **b,** The JMA catalog for ordinary earthquakes was used. Blue curve shows $M_c$ as a function of time. We used a moving event window of 300 events with a step of one event and attributed each calculated $b$-value to the origin time of the last event in each window (light-blue horizontal segment). One standard deviation of $M_c$ for each $M_c$-value is indicated by a light-blue vertical segment. Also included in **b** is the $M$-time diagram. Vertical line indicates the moment of the Mar. 11, 2011 $M9$ Tohoku earthquake.

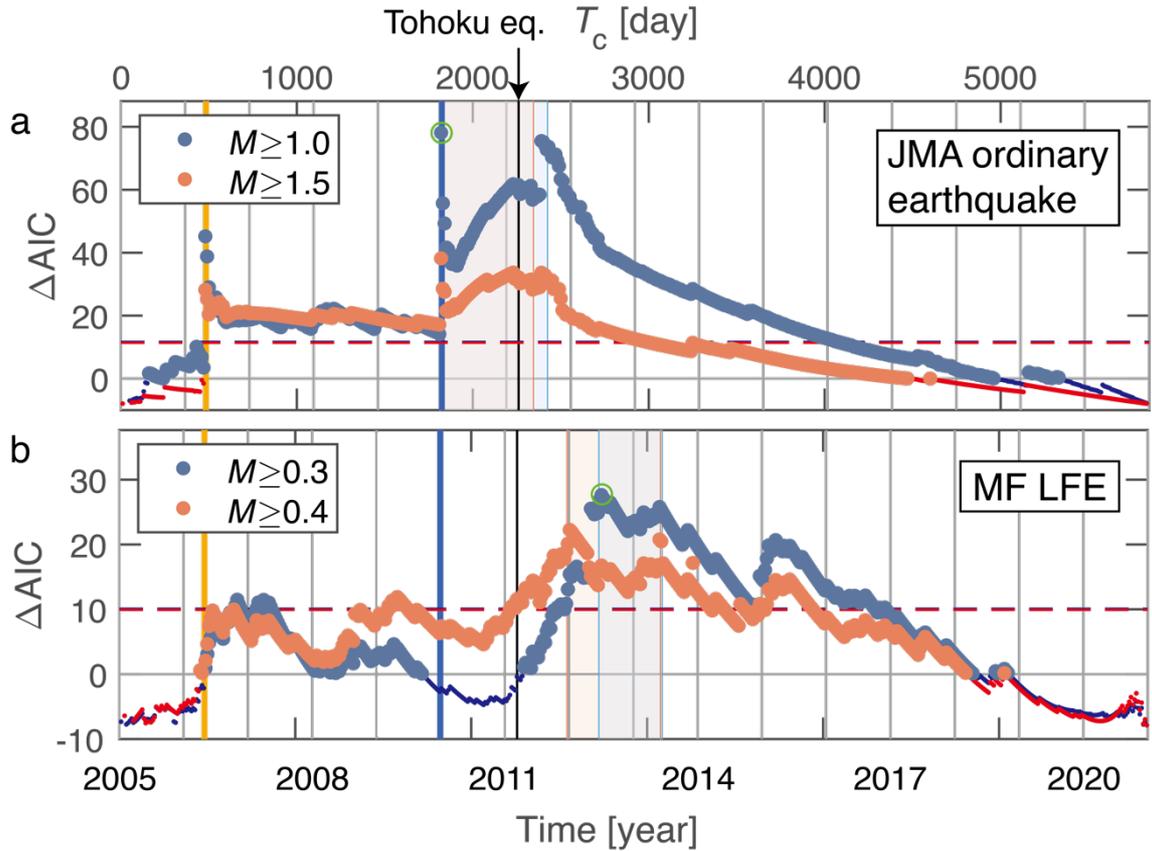

**Fig. 5.** First-order timeseries analysis. **a**, ΔAIC as a function of time (x-axis on the bottom side), taking $M_{th}$=1.0 (blue data) and $M_{th}$=1.5 (red data), where we used the JMA catalog of ordinary earthquakes. Grey vertical lines indicate Jan. 1 for 2006–2020. Also included in **a** for reference is $T_c$ (x-axis on the top side). Small points show that the model-fitting analysis did not converge when assuming the corresponding $T_c$. Orange and blue vertical lines indicate the moment of the 2006 and 2009 swarms, respectively. Horizontal dashed lines representing $2q$ for $M_{th}$=1.0 (blue) and 1.5 (red) overlap. Blue semitransparent region during $T_c$=1815-2425 days represents the change point's confidence interval of 68% for $M_{th}$=1.0. Red semitransparent region during $T_c$=1815-2345 days represents this interval for $M_{th}$=1.5. The ETAS fitting for the data point indicated by a green circle is shown in Fig. 6**a**,**b**. Vertical line indicates the moment of the Mar. 11, 2011 *M*9 Tohoku earthquake. **b**, Same as **a** for the MF catalog (*CC*>0.25) of LFEs. Change point's confidence interval is $T_c$=2725-3085 days for $M$≥0.3 (blue) and $T_c$=2545-3075 days for $M$≥0.4 (red). The ETAS fitting for the data point indicated by a green circle is shown in Fig. 6**c**,**d**.

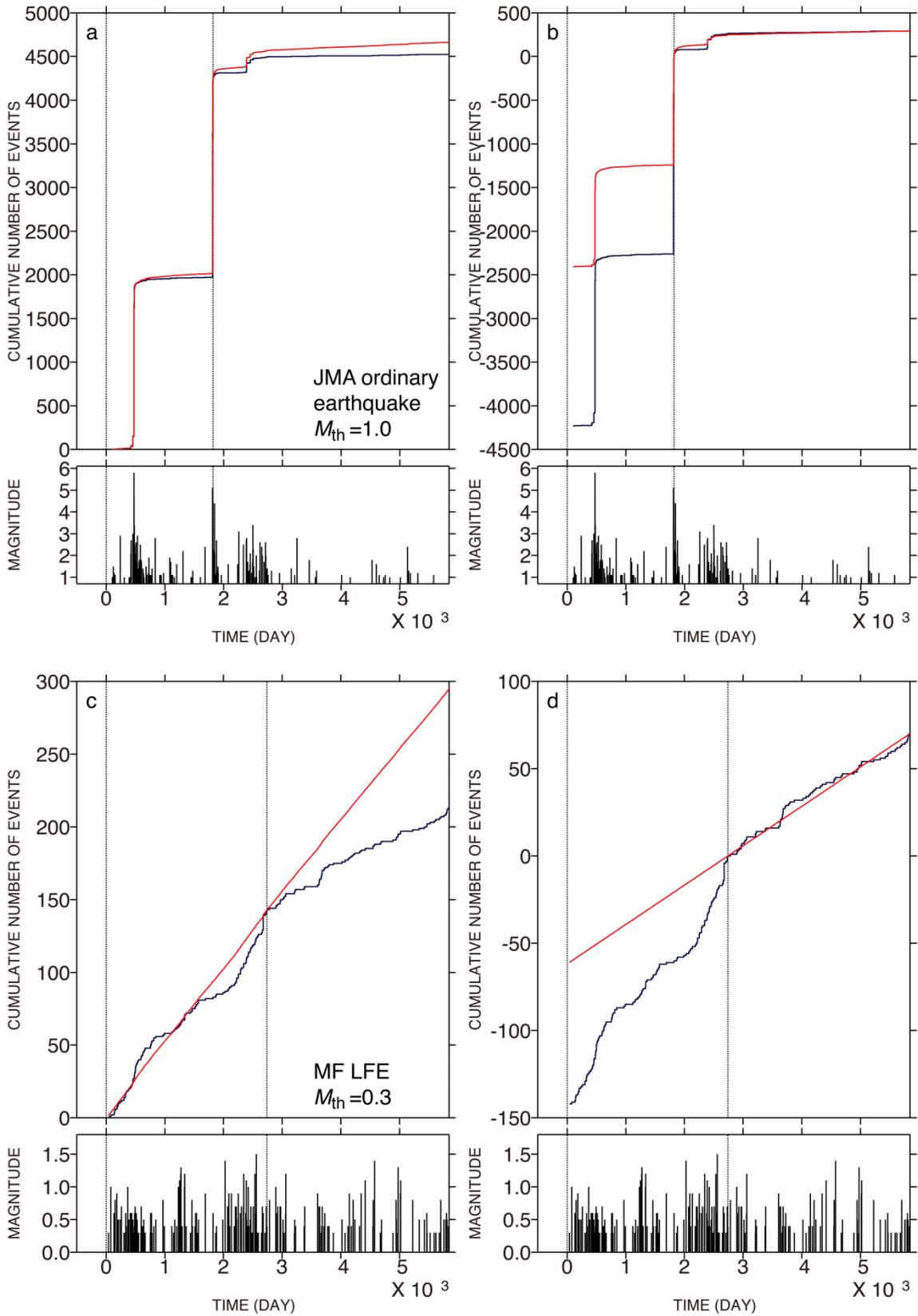

**Fig. 6.** Change point analysis for the first-order timeseries analysis. **a**, Cumulative function of $M \geq 1.0$ for ordinary earthquakes (black curve) is plotted against time, showing the standard (single) ETAS

fitting (red curve) in the interval from Jan. 2005 (first vertical line) to $T_c$=1820 days (Dec. 26, 2009) (second vertical line) and extrapolating until Dec. 2020. $T_c$=1820 days corresponds to $T_c$ indicated by the green circle in Fig. 5a. The set of parameters is θ=(μ, $K_0$, $c$, $α$, $p$)=(0.019, 0.032, 0.001, 0.058, 1.40). The extrapolation is the occurrence rate computed by using the standard (single) ETAS model with this θ. The $M$-time diagram is shown below the panel. b, As in a except for the ETAS fitting in the interval from $T_c$=1820 days (Dec. 26, 2009) (second vertical line) to Dec. 2020 and extrapolating until Jan. 2005 (first vertical line). The set of parameters is θ=(0.003, 0.024, 0.001, 0.000, 1.27). The extrapolation is the occurrence rate computed by using the standard (single) ETAS model with this θ. c, As in a except for LFEs ($M$≥0.3) and $T_c$=2740 days (Jul. 4, 2012). $T_c$=2740 days corresponds to $T_c$ indicated by the green circle in Fig. 5b. The set of parameters is θ=(0.046, 0.00035, 0.001, 0.00, 1.80). d, As in b except for LFEs ($M$≥0.3) and $T_c$=2740 days (Jul. 4, 2012). The set of parameters is (0.023, 0.00000, 0.001, 0.00, 1.58).

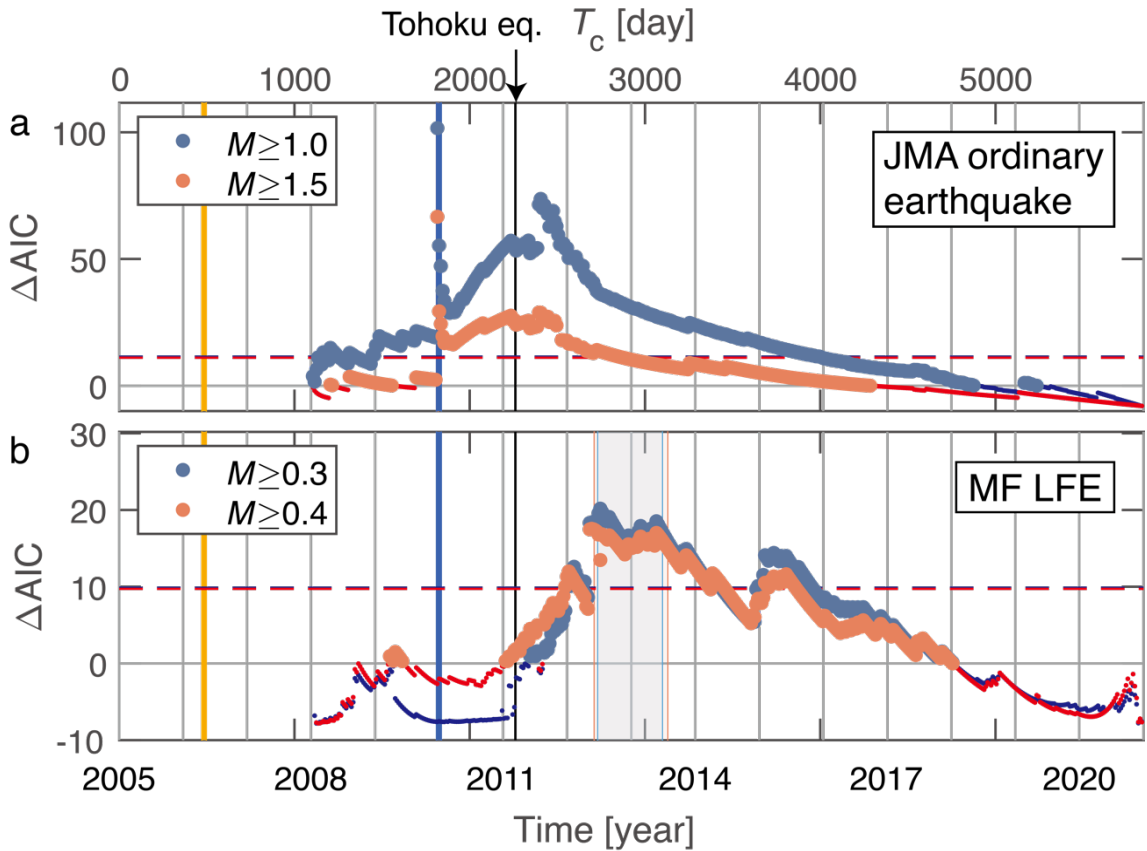

**Fig. 7.** Same as Fig. 5 for seismicity since Jan. 2008. For an explanation of symbols, lines, and regions, see the caption of Fig. 5. In **a**, the change point's confidence interval ($T_c$=1810-1820 days for $M_{th}$=1.0

and 1.5) overlaps with the blue vertical line indicating the moment of the 2009 swarm. In **b**, this interval is $T_c$=2730-3100 and 2710-3130 days for $M_{th}$=0.3 and 0.4, respectively.

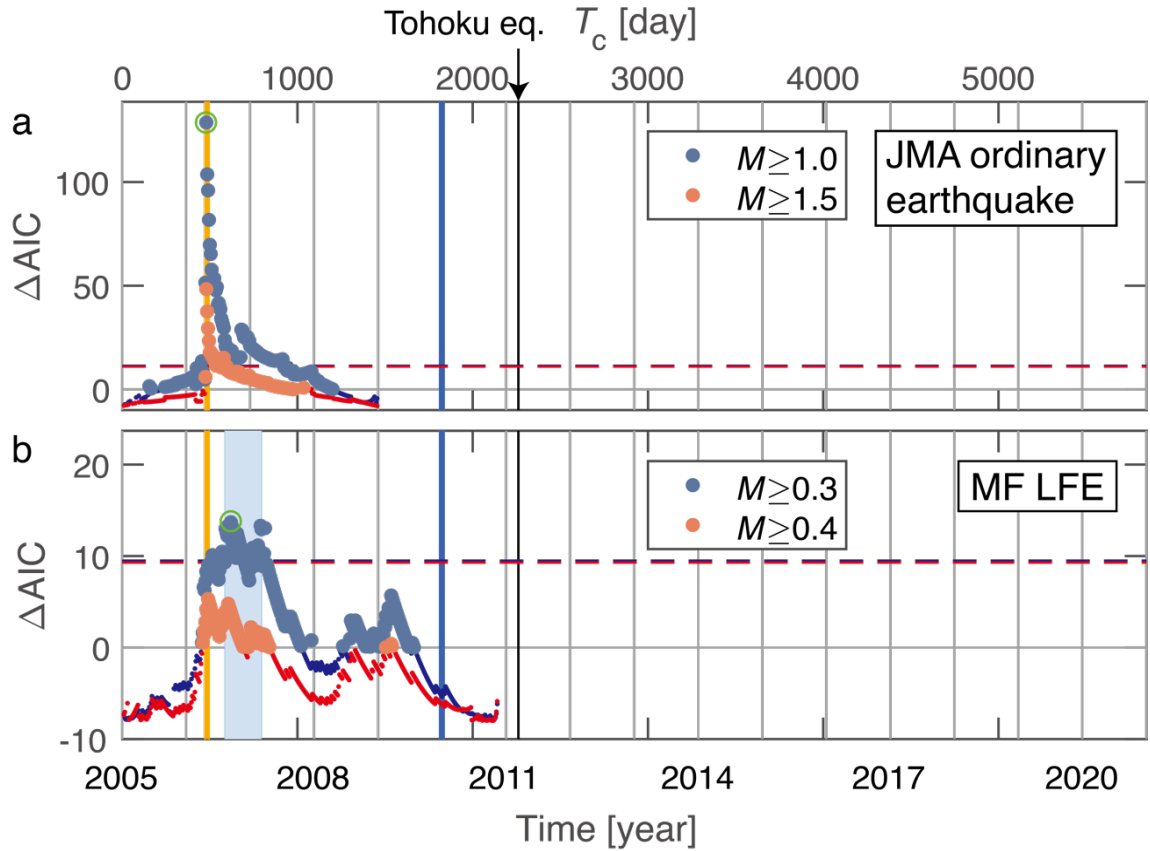

**Fig. 8.** Same as Fig. 5 for the second-order timeseries analysis. For an explanation of symbols, lines, and regions, see the caption of Fig. 5. In **a**, the change point's confidence interval ($T_c$=475-485 days for $M_{th}$=1.0 and 1.5) overlaps with the orange vertical line indicating the moment of the 2006 swarm. $T_c$=480 days with the largest ΔAIC, indicated by a green circle, is considered in Fig. 9**a,b**. In **b**, this interval is $T_c$=585-795 days for $M_{th}$=0.3. No change point's confidence interval for $M_{th}$=0.4 is shown because data points of ΔAIC for $M_{th}$=0.4 are below the horizontal red dashed line indicating $2q$, a hurdle to selection of the two-stage ETAS model. The ETAS fitting for the data point indicated by a green circle ($T_c$=620 days) for $M_{th}$=0.3 is shown in Fig. 9**c,d**.

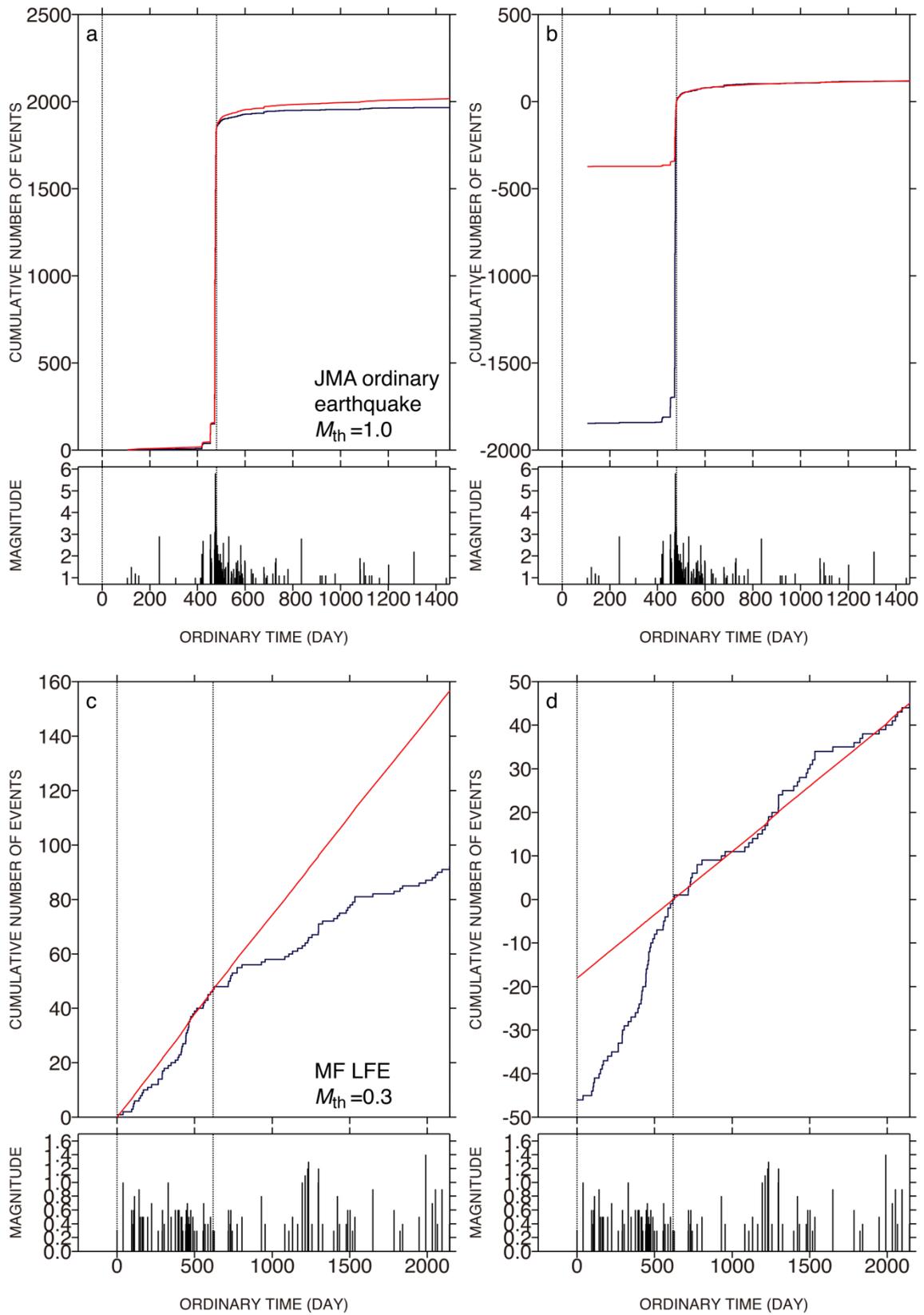

**Fig. 9.** Same as Fig. 6 for second-order timeseries analysis. **a**, Cumulative function of $M \geq 1.0$ for ordinary earthquakes (black curve) is plotted against time, showing the standard (single) ETAS fitting

(red curve) in the interval from Jan. 2005 (first vertical line) to $T_c$=480 days (second vertical line) and extrapolating until Dec. 2008. $T_c$=480 days corresponds to $T_c$ indicated by the green circle in Fig. 8**a**. The set of parameters are θ=(μ, $K_0$, $c$, α, $p$)=(0.024, 0.03, 0.001, 0.14, 1.47). **b**, As in **a** except for the ETAS fitting in the interval from $T_c$=480 days (second vertical line) to Dec. 2008. The set of parameters is θ=(μ, $K_0$, $c$, α, $p$)=(0.010, 0.07, 0.001, 0.34, 1.12). **c**, As in **a** except for LFEs with $M$≥0.3. The standard (single) ETAS model was fitted to the interval from Jan. 2005 (first vertical line) to $T_c$=620 days (second vertical line) and extrapolating until Dec. 2010. $T_c$=620 days corresponds to $T_c$ indicated by the green circle in Fig. 8**b**. The set of parameters is θ=(0.069, 0.00014, 0.001, 0.00, 1.91). **d**, As in **c** except that the standard (single) ETAS model was fitted to the interval from $T_c$=620 days (second vertical line) to Dec. 2010 and extrapolating until Jan. 2005 (first vertical line). The set of parameters is (0.029, 0.00000, 0.001, 5.69, 4.51).

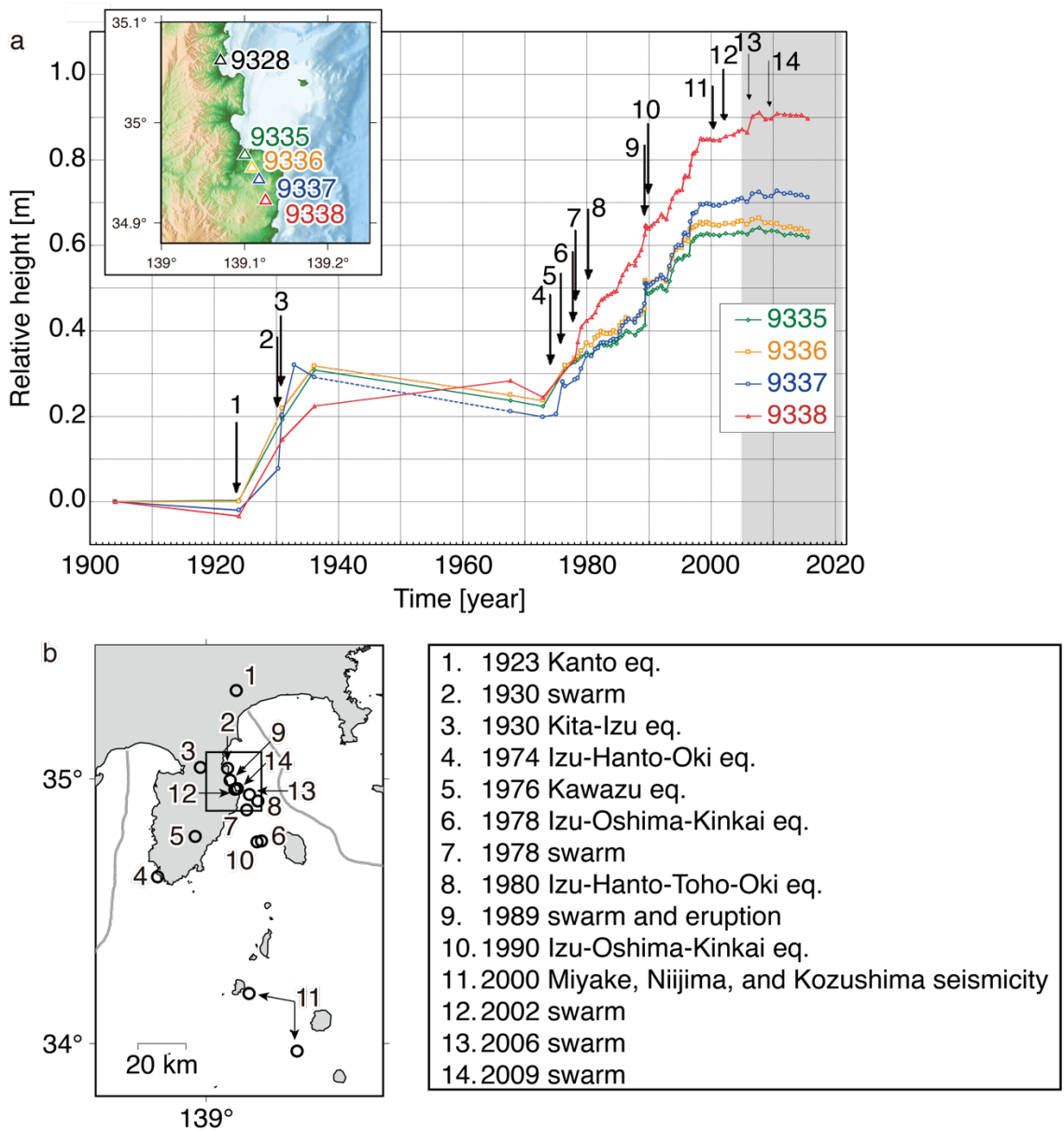

**Fig. 10.** Timeseries of vertical crustal movement. **a**, The leveling survey refers to benchmark 9328. Taking the year 1904 as the reference time, the results were obtained by the leveling survey at benchmarks 9335, 9336, 9337, and 9338. Data points and arrows indicating the 14 events were reproduced from GSI (2016). To directly compare between the long time period and a short time period that we are currently analyzing, a grey region indicating the study period (2005-2020) is included. Inset shows the locations of the benchmarks. **b**, Map showing the locations of the 14 seismic events, where the JMA catalog of ordinary earthquakes was used. The epicenters of the earthquakes (1, 3-6, 8, and 10) and the largest events in the swarms (2, 7, 9, and 11-14) are shown. For 11, there were two earthquakes with the largest magnitude *M*6.5, so the epicenters of these earthquakes are shown. Rectangle indicates the region of the inset of **a**. Gray curves indicate trough axes.

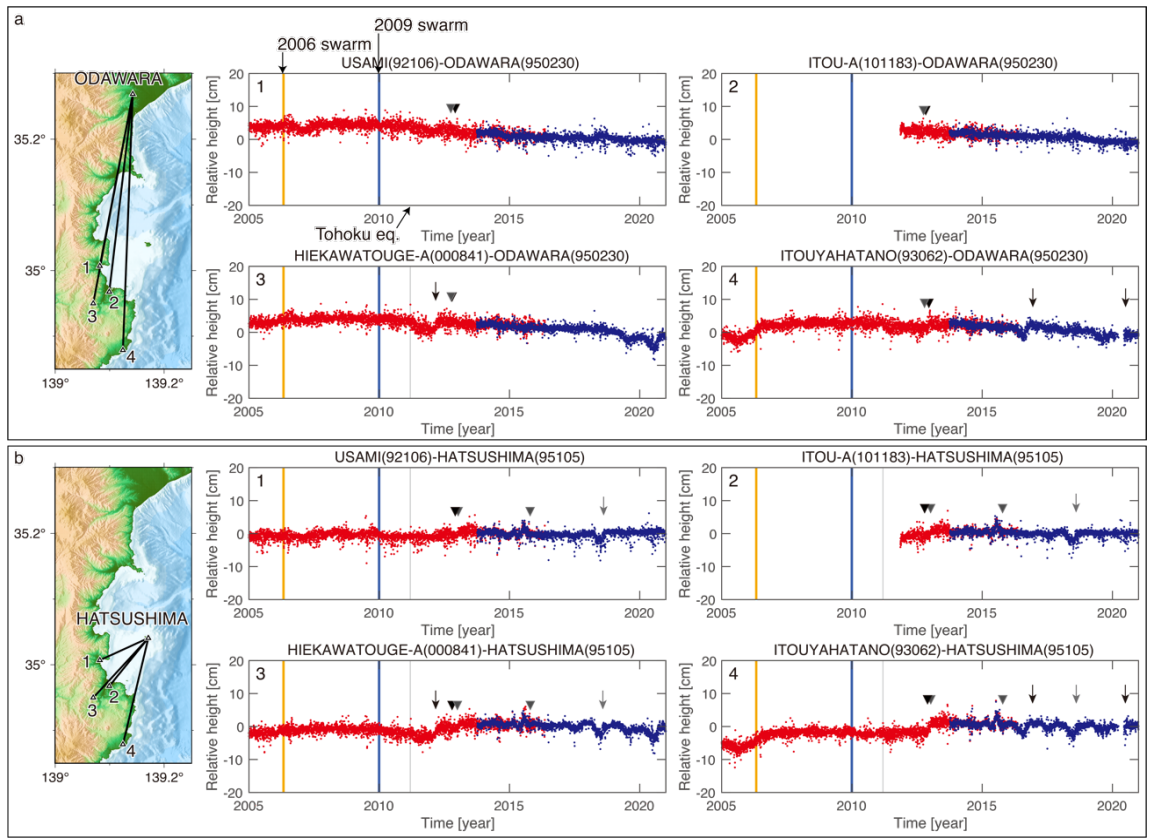

**Fig. 11.** Baseline changes. **a**, Left panel: Map showing stations (triangle) and baselines (segment). Reference station is ODAWARA, and stations 1, 2, 3, and 4 are USAMI, ITOU-A, HIEKAWATOUGE-A, and ITOUYAHATANO, respectively. Middle top panel: Relative heights of station 1 to the reference station as a function of time. Red and blue dots are F3 and F5 solutions, respectively. Orange and blue vertical lines indicate the moment of the 2006 and 2009 swarms, respectively. Right top panel: Same as the middle top panel for ITOU-A. Middle bottom panel: Same as the middle top panel for HIEKAWATOUGE-A. Right bottom panel: Same as the middle top panel for ITOUYAHATANO. Major maintenance: Triangle indicates the replacement of antennas at stations 1-4 (black) and the reference station (grey), and arrow indicates tree trimming around stations 1-4 (black) and the reference station (grey). **b**, Same as **a** for reference station HATSUSHIMA.

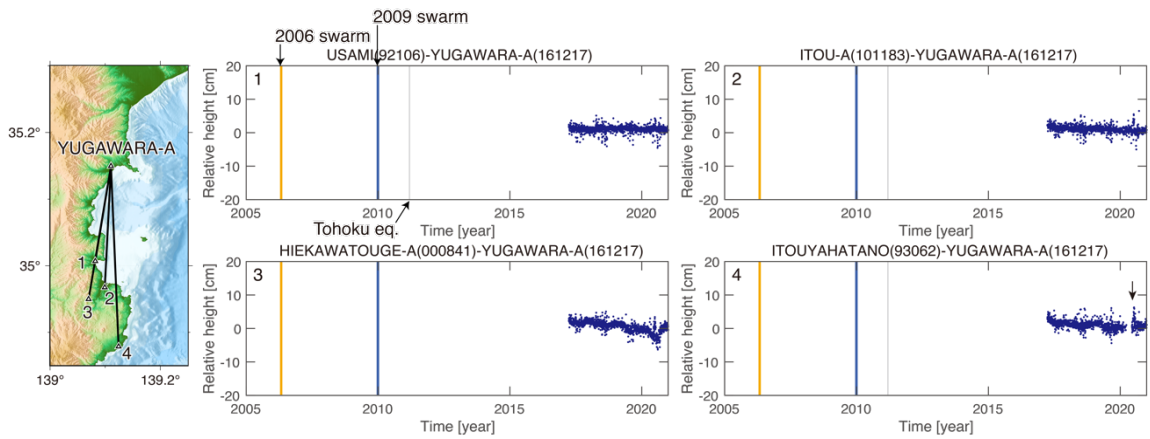

**Fig. 12.** Same as Fig. 11 for reference station YUGAWARA-A.

Supplementary Information

Changes in seismicity in a volcanically active region of the Izu Peninsula, Japan

K. Z. Nanjo, Y. Yukutake, T. Kumazawa

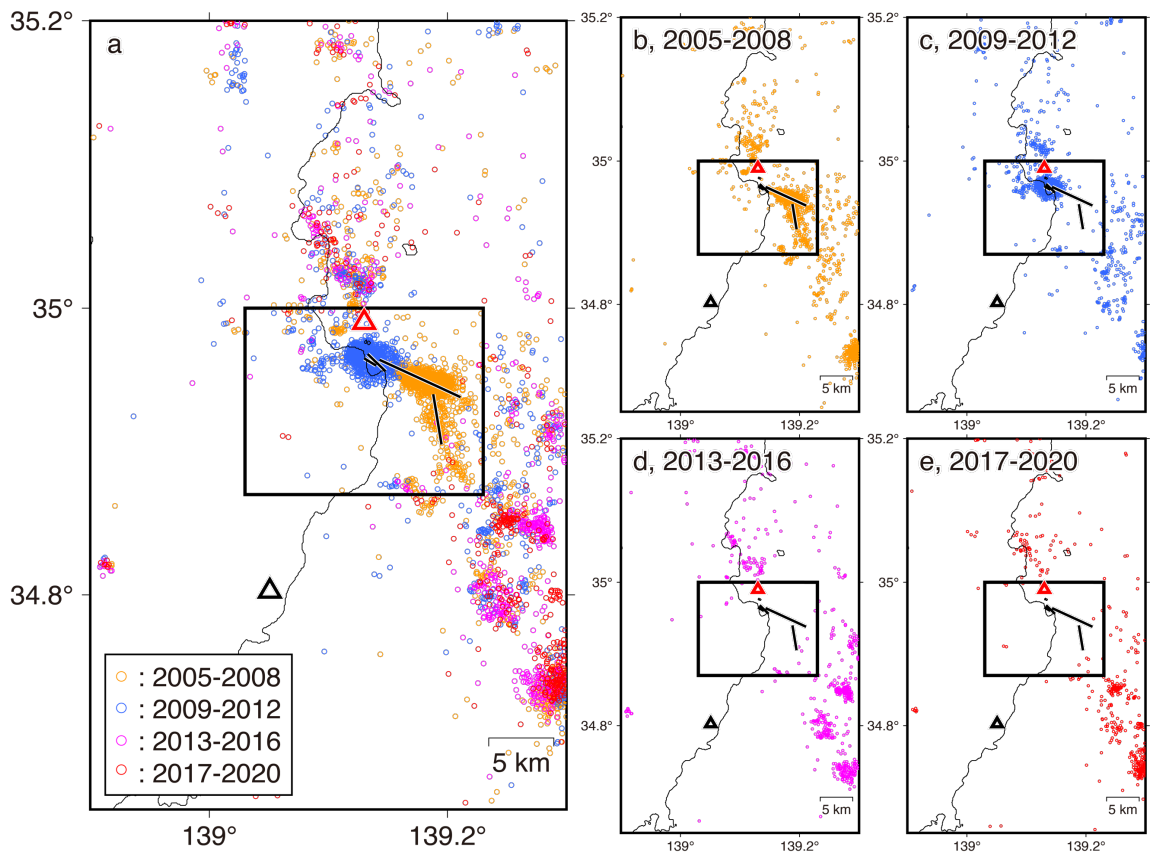

Supplementary Figure 1. Ordinary earthquakes colored by time. We used ordinary earthquakes ($M≥1$ and depth≤40 km) in the JMA catalog. **a**, Orange, blue, magenta, and red circles indicate earthquakes in 2005-2008, 2009-2012, 2013-2016, and 2017-2020, respectively. For red and black triangles, black rectangle, and black segments, see the caption of Fig. 1. **b**, As in **a**, but only earthquakes in 2005-2008 are shown. **c**, As in **a**, but only earthquakes in 2009-2012 are shown. **d**, As in **a**, but only earthquakes in 2013-2016 are shown. **e**, As in **a**, but only earthquakes in 2017-2020 are shown.

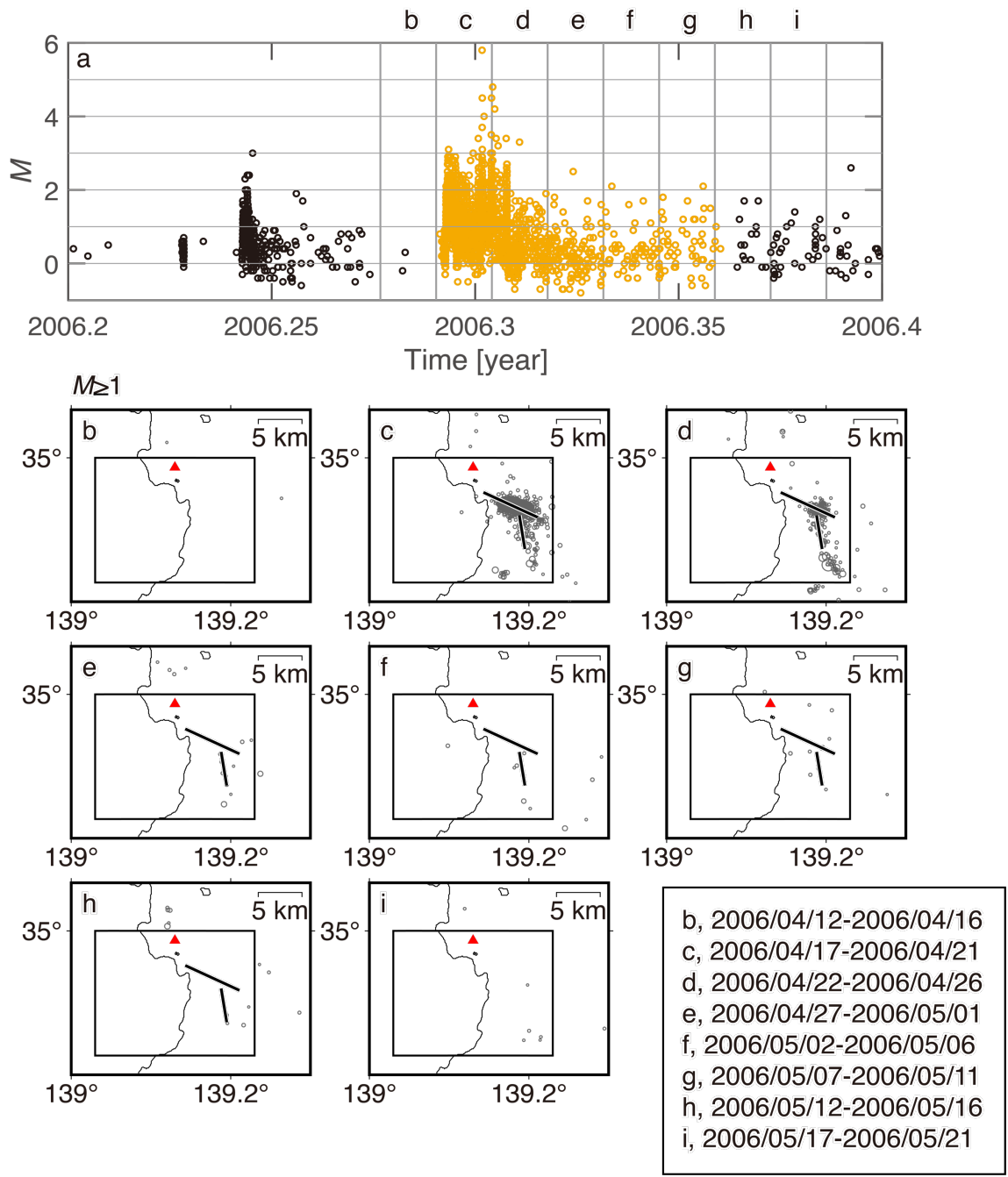

Supplementary Figure 2. Seismicity before, during, and after the 2006 swarm. We used ordinary earthquakes (depth≤40 km) in the JMA catalog. **a**, The *M*-time diagram for earthquakes (*M*≥-1) during the period 2006.2-2006.4 decimal year corresponding to Mar. 15-May 27, 2006 in the rectangular region shown in Fig. 1**b**. Orange circles indicate earthquakes in the major swarm in 2006 (Apr. 17-May 12, 2006) defined by JMA (2014). The periods indicated by b, c, …, i were considered in **b, c**, …, **i**. **b**, Spatial distribution of earthquakes (*M*≥1) during the period indicated by b (Apr. 12-16, 2006) in **a**. The rectangle region is the same as that shown in Fig. 1**b**. Red triangle indicates the Teishi Knoll.

**c**, **d**, …, **i**, Same as **b** for the periods indicated by c, d, …, i in **a**, respectively. A set of the two segments shown in **c**, **d**, …, **h** is the 2006-swarm source modeled by the GSI (2007).

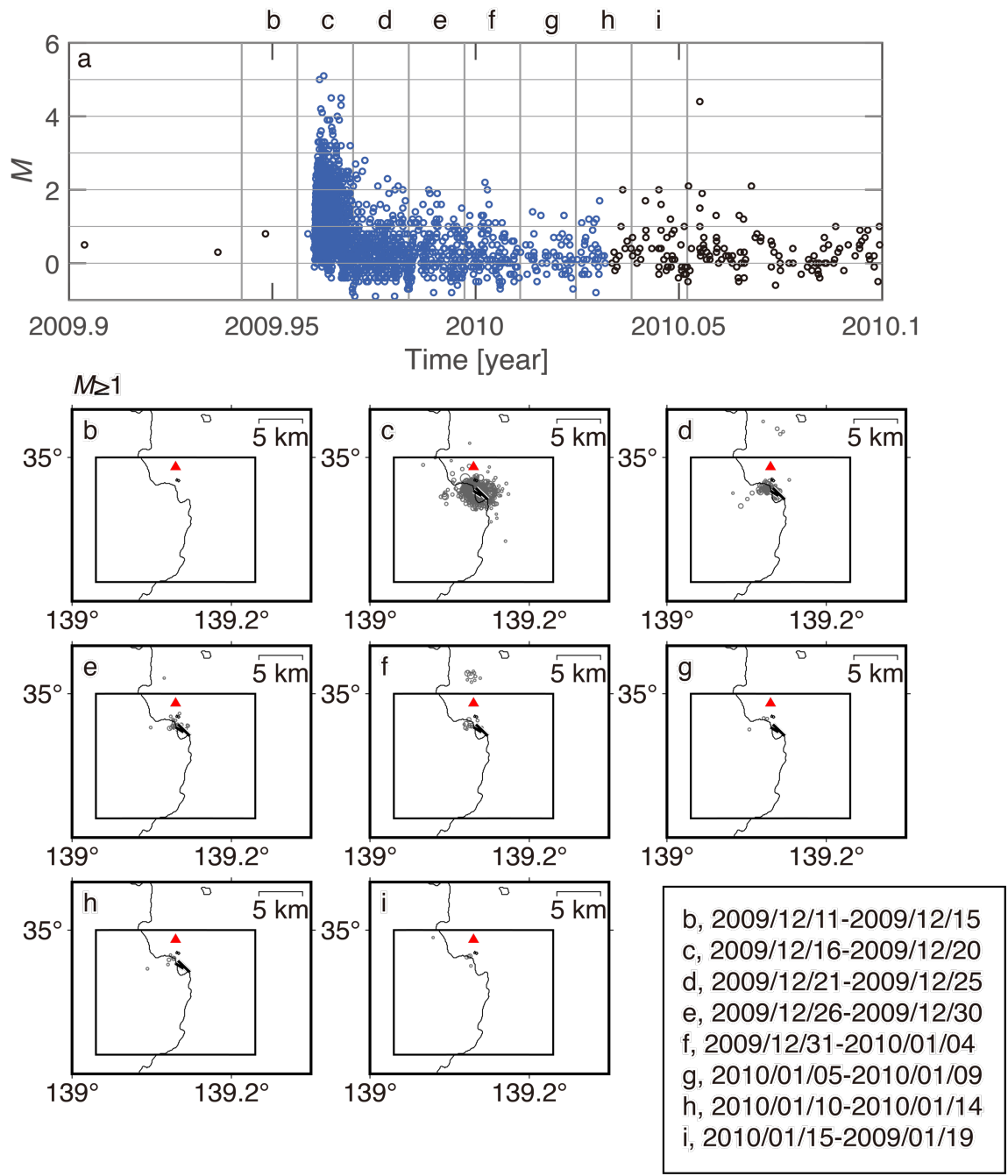

Supplementary Figure 3. Same as Supplementary Fig. 2 for the 2009 swarm. In **a**, blue circles indicate earthquakes in the major swarm in 2009 (Dec 16, 2009-Jan. 12, 2010) defined by JMA (2014). A set of the two lines shown in **c**, **d**, …, **h** is the 2009-swarm source modeled by the JMA (2010).

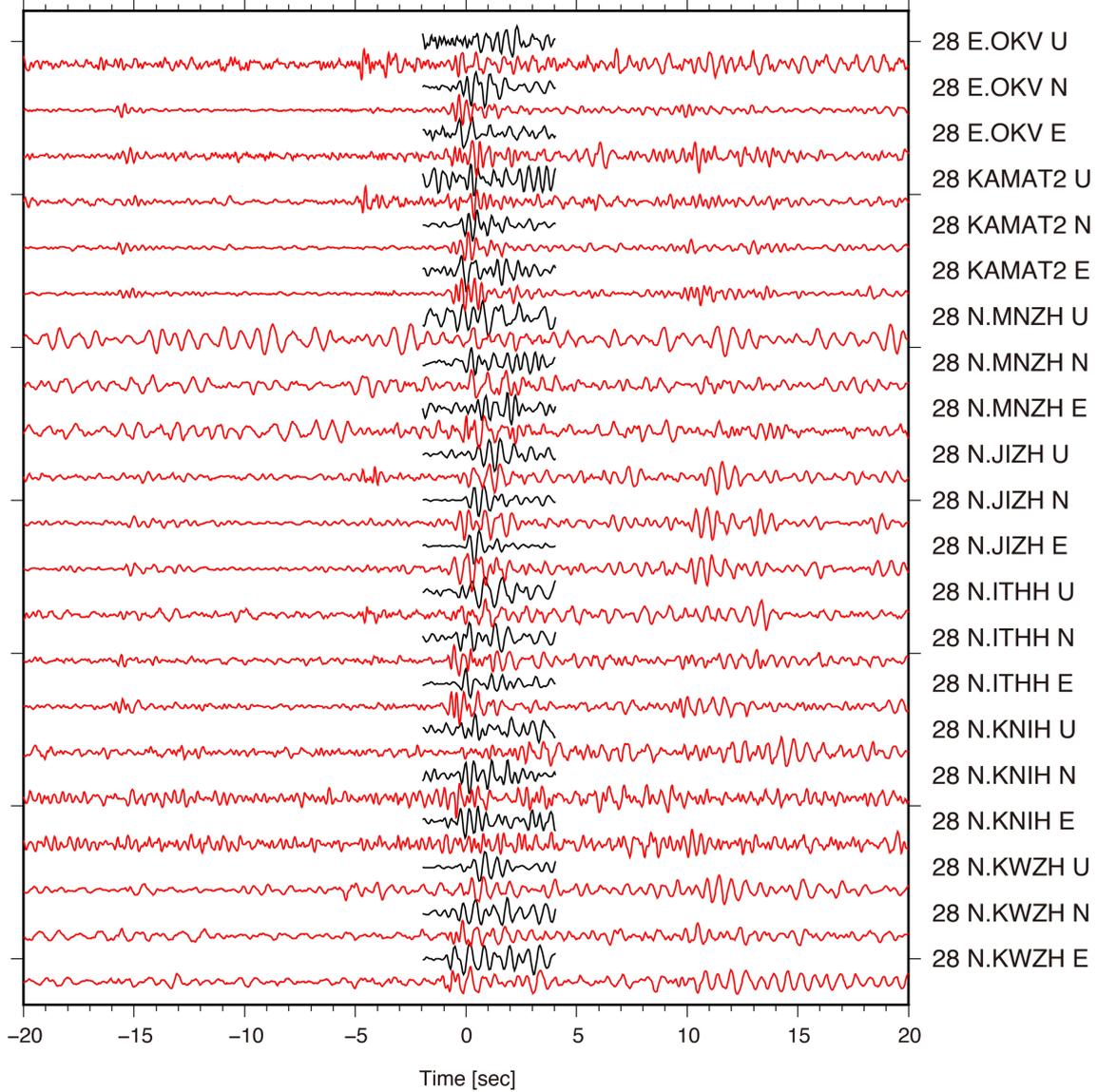

Supplementary Figure 4. Example of the MF method. Continuous waveforms (red) and their matched template waveforms (black) at each channel near the arrival times of the detected event. Event information is as follows. Time: Jan. 7, 2011, 20:15:55; assigned location: 139.12367°E, 34.93450°N, and depth 34.48 km; *M*=0.2; *CC*=0.393; and template event ID 28. Station names with three components (U, N, and E) used to detect the event are indicated on the right axis.